\newif\ifMNRAS
\MNRAStrue 

\PassOptionsToPackage{pdfpagelabels=false}{hyperref} 

\ifMNRAS
    \documentclass[fleqn,usenatbib,useAMS]{mnras}
    \usepackage{graphicx}
\else
    \documentclass[twocolumn]{aastex63}
\fi

\usepackage{acronym}
\usepackage{hyperref}
\usepackage{amsmath,amsfonts,amssymb}
\usepackage{mathrsfs}
\usepackage{mathtools}
\usepackage[utf8]{inputenc}

\usepackage[normalem]{ulem}
\usepackage[dvipsnames]{xcolor}
\usepackage{xspace}

\usepackage{pifont}

\newcommand{\tianshu}[1]{{ #1}}

\ifMNRAS
\title[Explosion Condition]{The Essential Character of the Neutrino Mechanism of Core-Collapse Supernova Explosions}

\author[Wang et al.]{
\href{0000-0002-0042-9873}{Tianshu Wang $^{1}$}\thanks{E-mail: tianshuw@princeton.edu}
\href{https://orcid.org/0000-0003-1938-9282}{David Vartanyan$^{2}$},
\href{https://orcid.org/0000-0002-3099-5024}{Adam Burrows$^{1}$},
\href{https://orcid.org/0000-0001-5939-5957}{Matthew S. B. Coleman$^{1}$}
\\
$^{1}$Department of Astrophysical Sciences, 4 Ivy Lane, Princeton University, Princeton, NJ 08544, USA\\
$^{2}$Department of Astronomy, University of California, Berkeley, CA  94720, USA\\
}

\date{Accepted XXX. Received YYY; in original form ZZZ}

\pubyear{2022}

\else
\shorttitle{condition}
\shortauthors{Wang et al.}
\fi

\begin{document}
\label{firstpage}
\pagerange{\pageref{firstpage}--\pageref{lastpage}}
\maketitle
\ifMNRAS
\else
\title{}

\affiliation{Department of Astrophysical Sciences, Princeton, NJ 08544, USA}
\fi

\begin{abstract}
Calibrating with detailed 2D core-collapse supernova simulations, we derive a simple core-collapse supernova explosion condition based solely upon the terminal density profiles of state-of-the-art stellar evolution calculations of the progenitor massive stars.  This condition captures the vast majority of the behavior of the one hundred 2D state-of-the-art models we performed to gauge its usefulness. The goal is to predict, without resort to detailed simulation, the explodability of a given massive star.  We find that the simple maximum fractional ram pressure jump discriminant we define works well $\sim$90\% of the time and we speculate on the origin of the few false positives and false negatives we witness. The maximum ram pressure jump generally occurs at the time of accretion of the silicon/oxygen interface, but not always. Our results depend upon the fidelity with which the current implementation of our code F{\sc{ornax}} adheres to Nature and issues concerning the neutrino-matter interaction, the nuclear equation of state, the possible effects of neutrino oscillations, grid resolution, the possible role of rotation and magnetic fields, and the accuracy of the numerical algorithms employed remain to be resolved. Nevertheless, the explodability condition we obtain is simple to implement, shows promise that it might be further generalized while still employing data from only the unstable Chandrasekhar progenitors, and is a more credible and robust simple explosion predictor than can currently be found in the literature. 
\end{abstract} 

\ifMNRAS
    \begin{keywords}
    stars - supernovae - general
    \end{keywords}
\else
    \keywords{
    stars - supernovae - general }
\fi

\section{Introduction}
\label{sec:int}






The astrophysics of core-collapse supernova (CCSN) explosions has achieved a significant degree of sophistication since the pioneering work of \citet{1966ApJ...143..626C}.  In that paper, the connection between the neutrino agency of explosion and core mantle heating by the neutrinos that are copiously produced due to core dynamical compression upon achieving the Chandrasekhar instability at the termination of stellar evolution was first hypothesized. However, the neutrino physics and transport, the initial stellar model, and the nuclear equation of state (EOS) were all quite primitive. \citet{1966CaJPh..44.2553A} improved upon all these factors, but did not find a robust explosion.  In the interim, the neutral current and $\tau$ neutrinos ($\nu_{\tau}$) were discovered and the nuclear EOS was made much more sophisticated, a process which continues to this day. 

It was not until the work of \citet{wilson1985} and \citet{1985ApJ...295...14B} that the ``delayed" neutrino-driven mechanism was introduced. In this model, the stalled shock generated upon core bounce after collapse and infall is reenergized after a post-bounce delay of hundreds of milliseconds by neutrino heating behind the shock in the so-called ``gain region."  The latter region of low, but importantly non-zero, neutrino absorption optical depth lies between the electron-type ($\nu_e$) and anti-electron  ($\bar{\nu}_e$) neutrinospheres and the shock. This model in broad outline is our current paradigm, but as it was originally conceived has many limitations. It was a spherical 1D model that did not explode without a boost in the $\nu_e$ and $\bar{\nu}_e$ neutrino luminosities of nearly $\sim$25\% due to inner core convection driven by fiat and a mixing-length prescription by a ``neutron-finger" (aka ``salt-finger") instability.  Such an instability, however, is not seen nor expected in modern proto-neutron-star (PNS) simulations \citep{bruenn_dineva}.  Lepton-driven Ledoux convection in the PNS is in fact seen in modern multi-D simulations \citep{2006ApJ...645..534D,nagakura2020}, but does not boost the $\nu_e$ and $\bar{\nu}_e$ neutrino luminosities by the required amount and is not a central factor in reenergizing the shock in modern theory.  The central factor in driving a stalled shock into explosion are now thought to be neutrino energy deposition/heating in the gain region, aided most of the time by neutrino-driven convection \citep{herant1994}.  The latter contributes a turbulent stress \citep{bhf1995} behind the shock and increases the dwell time in the gain region \citep{murphy2008} that facilitates net neutrino heating, with the added stress likely more important than the increased dwell time.  This overall picture is now strongly buttressed by the modern concordance in the results of many recent state-of-the-art 3D simulations \citep{2015ApJ...807L..31L,jms2016, roberts2016,muller2017,oconnor_couch2018b,vartanyan2018b,ott2018_rel,summa2018,glas2019,burrows_2019,nagakura2019b,burrows2020,2020ApJ...896..102K,bollig2021,2021ApJ...916L...5V,2021Natur.589...29B}. However, despite this palpable progress, it has proven difficult to convert this qualitative picture into a quantitative and predictive condition without these sophisticated 3D radiation/hydrodynamic simulations.  There have, however, been numerous attempts to do so, all of which are lacking in some important aspects.

The first attempt at this was the luminosity ($L_{\nu}$)/mass-accretion rate ($\dot{M}$) critical condition of \citet{burrowsgoshy1993}.  These authors noted that above a critical $L_{\nu}$/$\dot{M}$ curve 
\tianshu{no hydrostatic structure, including the core and the mantle interior to the shock, can be found.}  Increasing the luminosity above this curve led to a bifurcation, and did not merely increase the stalled shock radius quasi-statically. Furthermore, these authors speculated that the imaginary part of the frequency of the oscillation of the fundamental radial mode of shock pulsation would change sign into instability in traversing this curve.  That pulsation timescale is roughly $\frac{\Delta {M_g}}{\dot{M}}$, where $\Delta{M_g}$ is the mass in the gain region.  This is also roughly the advection time between the shock and the inner core in accretion flow. The supernova explosion solution itself was a dynamical transition between two steady eigenfunction solutions $-$ the quasi-static accretion phase and a steady-state neutrino-heated wind phase \footnote{This bifurcation was identified as a ``dual-cusp catastrophe" \citep{burrows2013}, but that suggestion has not been further developed.}. In any case, though there seems to be a core of truth in the concept of a critical curve, there was no analytic or physics-based derivation of the curve itself. Such requires a sophisticated model of the luminosity, and the latter requires radiative transfer, neutrino-matter microphysics, and a time-dependent model employing PDEs $-$ in short, a detailed numerical model. Therefore, though the critical condition remains a robust metric for determining explosion, it requires simulation results and is not an ab-initio predictor of explosion. Nevertheless, the concept of the crossing of a critical curve upon reaching the explosive state has gained traction in the literature, with numerous papers attempting improvements.

For example, \citet{summa2016} expanded this criterion to include multi-dimensional turbulence, explicit reference to the proto-neutron star mass, and an estimate of the gain radius. \citet{summa2018} and \citet{janka2012} (see also \citealt{jms2016}) generalized this prescription to ostensibly include some of the the effects of rotation. \citet{2012ApJ...746..106P} and \citet{2018MNRAS.481.3293R} introduced the ``antesonic condition," deriving from a simple model a relationship between the ratio of the isothermal sound and the escape speeds behind the stalled shock above which no steady state solution exists. They later extended this to include rotation and turbulence and factored in the adiabatic sound speed \citep{2021MNRAS.502.4125R}. \citet{2017ApJ...834..183M} introduced the ``Force Explosion" condition \cite{2021arXiv211010173G}, which uses two radiation parameters, one comparing the neutrino power deposited in the gain region with the accretion power and the second explicitly factoring in the neutrino optical depth in the accreted matter near the neutron star surface.  Using this approach, they rederive the antesonic and critical conditions and attempt a more general perspective. 

Inspired by the original critical condition concept, but wanting to produce a condition from which one can predict the outcome of collapse from the progenitor density profile alone (i.e., determine ``explodability"), \citet{2016ApJ...818..124E} derived a curve in $\mu_4$ and $\mu_4 \times M_4$ space, where $\mu_4$ is a measure of the slope of the mass density around the jump (frequently near the silicon/oxygen [Si/O] interface) in specific entropy to a value of four (per baryon per Boltzmann's constant) and $M_4$ is the interior mass at that jump. There is a corresponding jump in density at that same approximate position and we will return to its importance in the body of the paper. Only the initial progenitor mass density profile is employed by \citet{2016ApJ...818..124E} to determine explodability.  Calibrating on their estimates of the explosion energy and progenitor mass of the Crab and SN1987A supernovae, they derive a critical curve above which models would explode. In their construction, $\mu_4$ and $\mu_4 \times M_4$ are related indirectly to $L_{\nu}$ and $\dot{M}$, so the heritage in \citet{burrowsgoshy1993} is manifest.  However, as
\citet{2016MNRAS.460..742M} suggest and we advance below, the \citet{ 2016ApJ...818..124E} critical curve doesn't in fact seem to be a robust index of explodability.

Quantities which have proven important in characterizing the potential explodability of the quasi-stationary shock through which matter is accreting onto the PNS core and behind which neutrinos are heating that same matter are the advection timescale, $t_{adv}$, and the heating timescale, $t_{heat}$ \citep{Thompson_2000,murphy2008}.  In some senses, it is the competition between these quantities that determines instability to explosion and the potential for a dynamical transition from the steady state.  Sample definitions of these quantities are:

\begin{equation}
    t_{adv} (\sim t_{dwell}) \approx \frac{\Delta{M_g}}{\dot{M}}\, ,
\end{equation}
where $\dot{M}$ is the accretion rate and $\Delta{M_{g}}$ is the gain region mass,
and
\begin{equation}
    t_{heat} = \frac{\lvert{E_{g}}\rvert}{\dot{Q}_{heat}}\, ,
\end{equation}
where $E_{g}$ could be either the total energy, the gravitational energy, or thermal energy in the gain region and $\dot{Q}_{heat}$ is the neutrino power deposition rate in that same region. Equations for these quantities have varied in the literature
and have not been made rigorous.  However, if in a general sense $t_{adv}$ is larger than $t_{heat}$, then neutrino power deposition can heat the gain region before advection through this region brings the accreting matter into the inner core interior to the gain radius where there is net cooling.  It is this cooling region just above the approximate positions of the $\nu_e$ and $\bar{\nu}_e$ neutrinospheres that \tianshu{undermines pressure support for the outer mantle matter and} imperils explosion. A quick traverse through the gain region undermines explosion, while a long dwell time might facilitate it.

However, the astrophysics of this is a bit more subtle.  Even if $t_{adv}$ were infinite, continued neutrino heating in the gain region would quickly saturate to an equilibrium between heating and cooling, the latter due to neutrino emission via $e^-$ and $e^+$ capture on free nucleons. This cooling rate is roughly proportional to the temperature ($T$) to the sixth power, and given the roughly inverse-square neutrino flux dilution, the temperature would equilibrate to a $\frac{1}{r^{1/3}}$ radial profile.
In trying to predict/explain the solar wind,  \citet{1958ApJ...128..664P} showed that an ideal-gas atmosphere with a temperature profile shallower than $\frac{1}{r}$ could not maintain hydrostatic equilibrium unless there was a finite pressure at infinity.
The absence of such a bounding pressure implied to \citet{1958ApJ...128..664P} instability to the solar wind.  In our case, there is a bounding pressure due to the accretion ram.  Then, the question becomes: given the neutrino heating rate (which leads to a temperature profile) is that ram pressure low enough to lead to instability $-$ to an explosion?  Since the heating rate $\dot{Q}_{heat}$ depends upon the $\nu_e$ and $\bar{\nu}_e$ luminosities (times roughly the mean neutrino optical depth, $\tau_{\nu}$, in the gain region, of order $\sim$0.1) and the ram pressure depends upon $\dot{M}$ at the shock, comparing $t_{adv}$ and $t_{heat}$ to derive a condition for explosion is roughly equivalent to the classic critical condition pitting $L_{\nu}$ and $\dot{M}$ \citep{burrows2013}.  

These considerations are qualitatively useful and provide insight into the nature of the neutrino-driven mechanism of CCSN explosion, but they are not predictive.  To determine which massive-star progenitors explode and to derive a working and credible explosion criterion one must have a detailed model for the luminosities and heating, for the equation of state, for actual neutrino-matter microphysics, and for the turbulence. Moreover, one must be able to gauge the feedback influence of the accretion rate on the luminosities and neutrino spectra themselves.  Hence, short of detailed radiation/hydrodynamic supernova simulations, even informed analytic treatments are likely to fall short\footnote{Though, one may be able to find interesting potential correlations between observables \citep{2016MNRAS.460..742M}.}.  Therefore, while serving to highlight important aspects that factor into explodability, all published approximate criteria for explodability fail to be usefully predictive.  

The compactness parameter of the initial progenitor density profile \citep{2011ApJ...730...70O, oconnor2013} has perhaps been the most misapplied explosion metric. It is defined as \citep{2011ApJ...730...70O}:
\begin{equation}
\xi_M= \frac{M/M_{\odot}}{(R(M)/1000\, \mathrm{km})}\, ,
\end{equation} 
where $M$ denotes the interior mass at which the compactness parameter is evaluated \footnote{originally defined at 2.5 M$_{\odot}$, but our preferred value is 1.75 M$_{\odot}$)}.  $\xi_M$ has been used as a parameter below which explosions would occur and above which they would not. However, this dichotomy has been shown to be false, with both low and high compactness models exploding \citep{burrows2020,2021Natur.589...29B}\footnote{though, perhaps the highest $\xi_M$ models abort explosion and reimplode to black holes}. 

Moreover, within the context of the neutrino-driven paradigm, higher compactness seems necessary to achieve explosion energies in the $\sim 10^{51}$ ergs (1 Bethe) range \citep{2021Natur.589...29B}. It is only such models that have enough neutrino absorption optical depth behind the shock for a sufficiently long period of time to absorb sufficient energy during launch. Such models also have more core ejecta mass whose recombination energy is thereby usefully high.

The advantage of compactness is that it is an ab-initio, simple-to-calculate parameter that would not require running detailed core-collapse simulations. Moreover, there is a strong correlation of compactness with post-bounce neutrino luminosity and accretion rate and, for exploding models that birth neutron stars, there is a strong positive correlation with residue mass. If a black hole forms, the timescale for its formation is inversely related to compactness, \tianshu{since accretion rate increases with compactness}. But, compactness per se does not serve as a useful explosion discriminant.

Finally, for most progenitors the mass at which $\xi_M$ is calculated also encompasses the silicon/oxygen (Si/O) interface entropy and density jumps and, as has been shown in the past \citep{swbj16,2016ApJ...818..124E,burrows2018,vartanyan2018a,burrows_2019,2021Natur.589...29B,2021ApJ...916L...5V}, this jump is an important feature that can affect explodability. This topic will be more extensively discussed in this paper.   

Given the degree to which attempting to derive a simple explodability criterion in the past has been fraught, it is reasonable to cast a jaundiced eye on attempts to do so now, or in the near future.  There is still much to do with regards to the initial progenitor models, the neutrino-matter microphysics, numerical algorithms, and computational implementations.  We do not think that current sophisticated models have yet converged to either 1) the ultimate answer or 2) the definitive mapping between progenitor and outcome.  Nevertheless, there has of late been sufficient progress in the multi-dimensional modeling of collapse and explosion to take a fresh look, guided by results emerging from the more detailed computational models.  In this paper, we acknowledge the physical insights provided by the past literature on explodability criteria, but seek to derive a means to predict explodability given only the initial mass density profiles calibrated on a large set of two-dimensional (axisymmetric) CCSN simulations using our code F{\sc{ornax}} \citep{skinner2019}. A large suite of three-dimensional simulations with which to benchmark success is still too expensive and time-consuming.  We have seen that, though the details of 2D and 3D evolutions can be different, with very few exceptions when one of our F{\sc{ornax}} models explodes in 2D it does so in 3D, and vice versa.  Under this assumption, and with the presumption that our simulations are capturing much of the essential truth about explodability (our major and most fragile assumption), we in this paper take up the challenge to develop a predictive explosion condition.  

First, we assembled many of our recent 3D and 2D models and explored with them various possible explodability conditions and metrics using only the associated progenitor mass density profiles with which those models were inaugurated.  Settling on one measure (and allowing some exceptions), we then picked 100 massive-star progenitor models from the compilation of \citet{swbj16} and \citet{sukhbold2018} and performed 2D CCSN F{\sc{ornax}} simulations for each of them for many seconds post-bounce to determine whether they exploded by the criterion we arrived at during the earlier exploratory phase.  Indeed, we found that the explodability criterion we ``predicted" comported with the results for the 100 models, with few exceptions.  This gave us confidence in the relative usefulness of the simple approach to determining explodability we found.

In \S\ref{summary} of this paper, we first set the stage by summarizing various physical aspects of the neutrino mechanism rarely found written in one place. We feel that articulating, even in bulletized form, the essential elements of the modern neutrino mechanism of CCSN explosion will provide context and might prove useful. Then, in \S\ref{explore} we summarize the results of the exploratory set of 3D and 2D models and the systematic behavior of this ``training set" with progenitor characteristics.  The associated simulations are taken from \citet{burrows2020} and \citet{2021Natur.589...29B}.  In \S\ref{motivation}, we motivate our choice of central explosion condition using the exploratory models, and in \S\ref{suite} we present our results for the 100 2D models, discuss caveats, and advance our approximate approach to determining explodability.  We conclude in \S\ref{conclude} with a summary of what we have found and a listing of the strengths and potential weaknesses of our findings.

\section{Physics of the Neutrino Mechanism}
\label{summary}

We provide here some of the basic physical characteristics of the neutrino mechanism of core-collapse supernova explosions.  There has been no attempt to fully flesh-out these features, nor to justify our choices.  Rather, we hope that a list of clear declarative points, without the baggage of detail, might help orient the reader in advance of the more focussed discussion that follows. We admit that our compilation might be incomplete and is a bit idiosyncratic.  However, we know that this decades-old subject can at times seem confusing and less coherent than perceived by long-time practitioners.  In fact, there is significant agreement on many aspects of the theory that is emerging.

\begin{itemize}
    
    \item The onset of the core-collapse supernova explosion is a bifurcation between a quasi-steady PNS solution with a bounding accretion shock wave and a mantle ejection instability.  There are no intermediate stable solutions. The asymptotic state of the core is one with a progressively weakening neutrino-driven wind, whose onset depends upon the evolving pressures and ram pressures of the material of the inner explosion \tianshu{\citep{burrows1987_wind,burrowsgoshy1993}}. Should the latter be large, the wind phase will be delayed. The supernova is a dynamic transition solution between the quasi-steady accreting PNS and wind eigenvalue solutions.  The wind-phase mechanical power is lower than that experienced during the supernova phase due to the much larger absorbing mass and degree of neutrino power deposition during the latter.
    
    \item It is thought that the explosion condition might be derivable from a suitable pulsational analysis which would lead to eigenfrequencies whose fundamental (and dipole? \citealt{2013ApJ...765..110D}) modes would experience at explosion a change in the sign of their imaginary parts. This has not been definitively explored in a realistic context, but seems compelling.
    
    \item {Gravitational energy released due to collapse from a Chandrasekhar white dwarf to a proto-neutron star is the ultimate source of the explosion energy\tianshu{\footnote{Here and below, when we refer to the explosion energy, we always mean the sum of all its contributions: the internal, recombination, kinetic, gravitational, and total binding energy of the off-grid stellar envelope (negative). The use by some of what is called the ``diagnostic energy" is misleading and can be confusing, and we recommend it never be invoked.}}. However, though the total binding energy of a neutron star is some multiple of 10$^{53}$ ergs, most of this energy is radiated by neutrinos over tens of seconds. Only a fraction is radiated during the launch and powering of the explosion and only about half of that is radiated in the $\nu_e$ and $\bar{\nu}_e$ neutrinos that dominate neutrino heating.  Therefore, the neutrino mechanism is not merely $\sim$1\% (or less) efficient, but closer to $\sim$5$-$10\% efficient \tianshu{\citep{2016ApJ...818..123B,bollig2021,2021Natur.589...29B}}.}
    
    \item Nuclear burning in the inner mantle shells of the progenitor contributes at most $\sim$5$-$10\% to the total explosion energy \tianshu{\footnote{The energy available from the burning of oxygen to $^{56}$Ni is $\sim$0.1 Bethes per 0.1 M$_{\odot}$ of oxygen. The amount of oxygen that can be burned to, for instance, $^{56}$Ni in the extended core generally scales with the compactness. The fraction of this oxygen that will explosively be burned scales with the neutrino-driven explosion energy. Hence, since we argue that the latter scales roughly with compactness, the total nuclear energy available very roughly scales with compactness. Therefore, the fractional contribution of nuclear energy to the total is very roughly constant. Since the measured amount of ejected $^{56}$Ni varies from $\sim$ a few$\times$10$^{-3}$ to $\sim$0.1 M$_{\odot}$, and the inferred CCSN explosion energies vary from $\sim$0.1 to $\sim$2 Bethes, one obtains the roughly $\sim$5$-$10\% quoted \citep{muller2017c}.}}.
    
    \item Unlike Type Ia thermonuclear explosions, the mechanism of core-collapse supernovae need not disassemble the residual core to leave nothing behind.  The bound neutron star early created in collapse quickly becomes too bound to disassemble. Rather, the ongoing and prodigious neutrino losses from the PNS ``neutrino star" that leaves a progressively more bound PNS transfers energy to its mantle and it is the mantle that explodes. It is this energy transfer from core to mantle and the latter's ejection that is the essence of the neutrino mechanism. During the launching of the shock wave much of the deposited neutrino energy is used to do work against gravity.
    
    \item Hence, the CCSN mechanism is an energy transducer from gravitational to core-thermal to neutrino to mantle-thermal to (mostly) kinetic energy.
    
    \item Furthermore, the gravitational energy of infall that is used to dissociate the nuclei accreted through the stalled shock is retrieved upon the recombination of the ejected nucleons.  This source (recall that the specific binding energy of nuclei is near $\sim$10 MeV/baryon) constitutes a large fraction of the total asymptotic supernova explosion energy.  
    
    \item Radiation pressure is not a significant factor in the neutrino mechanism of explosion.  The neutrino Eddington luminosity is near $\sim$10$^{55}$ erg s$^{-1}$,
    far in excess of even the peak $\nu_e$ luminosity at breakout (which itself lasts only $\sim$10 milliseconds).
    
    \item The cumulative energy deposited by neutrinos in the gain region and mantle interior to the stalled shock bears no relation to the explosion energetics.  In fact, the total mantle energy is negative at the onset of explosion and has been secularly decreasing during the delay to explosion. Afterwards, neutrino heating and recombination together combine over time after the delayed instability to achieve the positive asymptotic energy of the supernova.
    
    \item It is thought that, in broad outline, the energy scale of CCSN explosions is ultimately and approximately set by the total binding energy of the stellar mantle exterior to the Chandrasekhar core. \tianshu{This is analogous to the self regulation seen in stellar winds, for which the escape velocity from the star is roughly the asymptotic velocity of the wind at infinity \citep{2011Ap&SS.336..163V}.} This implies that the greater this binding energy, the greater the explosion energy, unless the progenitor does not explode (perhaps due to an over-large binding energy that can't be overcome by the available neutrino heating). This feedback concept is distantly analogous to that for a stellar wind, for which asymptotic speeds scale with surface escape speeds.  
    
    \item Neutrino heating of the mantle drives turbulent convection between the stalled shock and the PNS core.  This is akin to boiling water on a stove. The turbulent stress contributes an effective additional pressure that enlarges the gain region and increases the stalled shock radius.  The larger gain region absorbs slightly more neutrino energy and the larger shock radius puts some of the gain region mass lower in the gravitational potential well.  Thus less bound, this mass is more easily unbound.
    
    \item The stress tensor of the turbulence is anisotropic \citep{murphy2008,2013ApJ...771...52M}, with the radius-radius component in the radial direction larger than the other components.  Moreover, the effective $\gamma$ connecting turbulent energy with turbulent pressure is near $\sim$2 (compare with $\sim$4/3), making turbulence a more efficient means of realizing pressure with energy density than thermal gas.  
    
    \item $t_{adv}$ is also relevant to the strength of the turbulence, since a large value facilitates the growth of the seed perturbations in the pre-shock infalling matter to the non-linear amplitude regime before settling into the inner core by increasing the number of e-foldings \citep{foglizzo:07}.  The magnitude and spatial scales of the seed perturbations in the progenitor convective zones are not currently well-known, but can factor into the explosion details. Current CCSN simulations witness turbulent Mach numbers between $\sim$0.1 and $\sim$0.5.
    
    \item $\nu_{\mu}$ and $\nu_{\tau}$ neutrino emission constitutes $\sim$50\% of the total binding energy losses.  However, their contribution to heating in the gain region is subdominant.  Their major positive effect on explodability is indirectly by helping to drive the Kelvin-Helmholtz shrinkage of the core through their radiation, which leads to an increase in the temperatures of the $\nu_e$ and $\bar{\nu}_e$ neutrinospheres and thereby to a hardening of the emergent $\nu_e$ and $\bar{\nu}_e$ spectra. Such a hardening leads to higher neutrino absorption and heating cross sections in the gain region.
    
    \item The breaking of spherical symmetry enabled in 2D and 3D allows simultaneous accretion and explosion, the former maintaining neutrino accretion power that can continue to energize the latter.  This can't happen in 1D (spherical) and is an important reason, along with the modeling of turbulence, CCSN simulations need to be done in multi-D.
    
\end{itemize}

\section{Training Set of 3D and 2D Core-Collapse Models}
\label{explore}

\begin{figure*}
    \centering
    \includegraphics[width=0.47\textwidth]{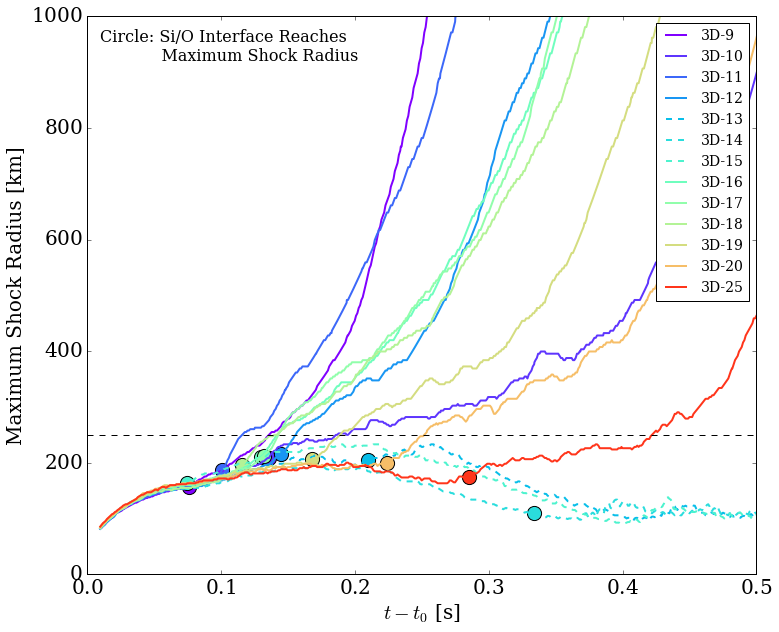}
    \includegraphics[width=0.47\textwidth]{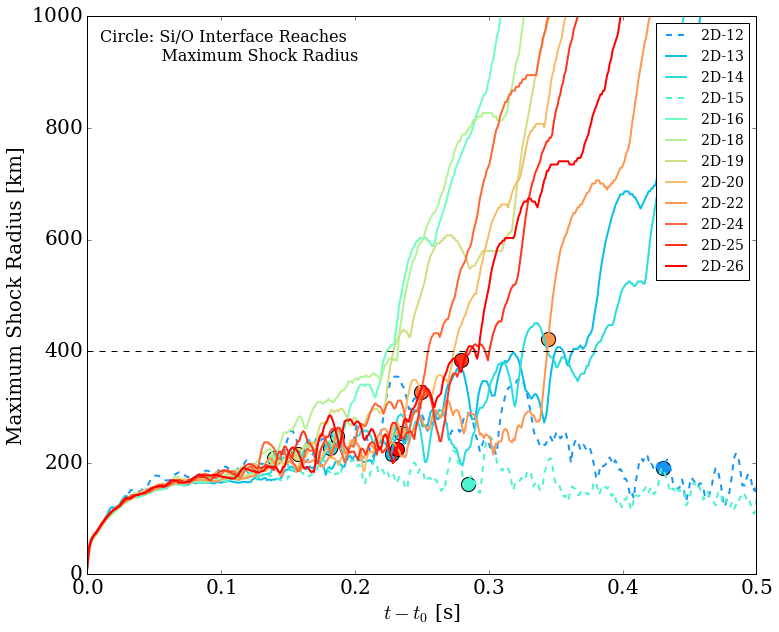}
    \caption{Evolution of the maximum shock radius. Left: 3D simulations from \protect\cite{burrows2020}. Right: 2D simulations from \protect\cite{2021Natur.589...29B}. The numbers after the ``3D" or ``2D" indicate the ZAMS mass of the progenitor in units of solar masses (M$_{\odot}$). Solid lines are exploding models, while dashed lines are non-exploding ones. The dot on each line marks the time when the maximum shock position encounters the Si/O interface. From the 3D simulations, we see that the maximum shock radius in each model evolves in a very similar fashion until the Si/O interface hits the shock. Despite experiencing stronger oscillations, the 2D simulations evince the same behavior. The dashed horizontal lines denote 250 km on left panels (re 3D) and 400 km on right panels (re 2D).}
    \label{fig:rshock}
\end{figure*}

As a prelude to this study, we investigated a set of legacy 3D \citep{burrows2020} and 2D \citep{2021Natur.589...29B} F{\sc{ornax}} simulations\footnote{These 3D and 2D studies were each done for a different set of progenitors and the associated papers describe the numerical and computational setups.} The early evolution of the maximum shock radius after bounce for these models is given in Figure \ref{fig:rshock}. The solid lines are for models that exploded and the dashed lines are for models that did not.  The filled circles indicate the time when the Si/O interface (or the interface with a large density drop) accretes through this maximum shock radius.  As is clear, for many models the time of explosion is roughly when (or near when) this occurs, and if a progenitor from this set explodes it generally does so after $\sim$70 to $\sim$350 milliseconds (ms) after bounce and the inauguration (at ``$t = 0$") of the quasi-steady delay phase.  Note that the models that don't explode (at least among these simulated models) have a low to intermediate ZAMS mass and that this model suite does not include ZAMS masses above 26 M$_{\odot}$.  We leave more massive progenitors to a future study. \tianshu{\footnote{Discussion on why progenitors in a certain mass range are less explosive can be found in Section \ref{suite}.}}

\begin{figure*}
    \centering
    \includegraphics[width=0.47\textwidth]{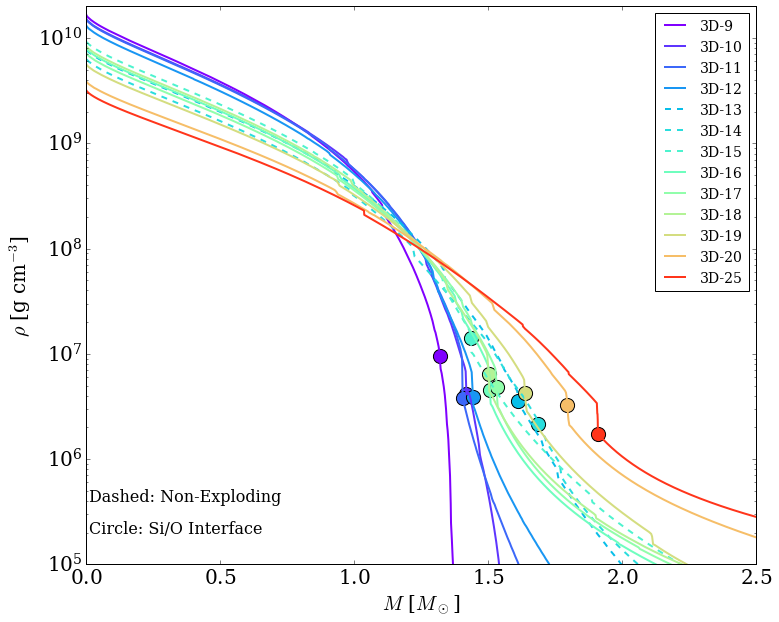}
    \includegraphics[width=0.47\textwidth]{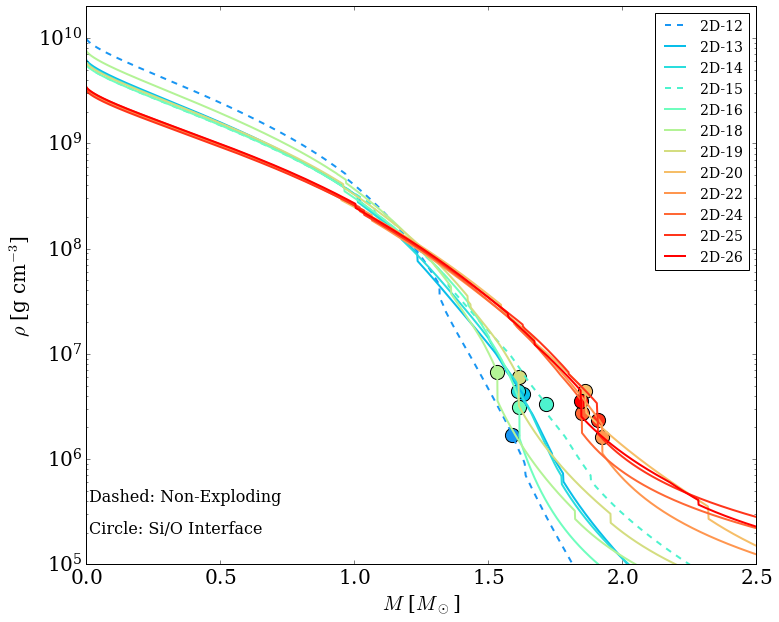}
    \caption{The $\rho$ (in gm cm$^{-3}$) versus interior mass ($M$, in M$_{\odot}$) profiles of the progenitors. Left: 3D simulations published in \protect\cite{burrows2020}. All these progenitors are taken from \protect\cite{swbj16}, except for the 25 $M_\odot$ model, which is taken from \protect\cite{sukhbold2018}. Right: 2D simulations published in \protect\cite{2021Natur.589...29B}.  The solid lines are exploding models, while the dashed lines are non-exploding models. The dot on each line marks the position of the Si/O interface defined via an abundance transition, which may slightly deviate from the start point of the associated density jump. Most non-exploding models generally have a weak density jump at the Si/O interface.} 
    \label{fig:rho-M}
\end{figure*}

The associated initial baryon mass density profiles of this model set are given in Figure \ref{fig:rho-M}, with the dots identifying the approximate positions of the Si/O interfaces.  It is our suggestion, shared by others, that, under the assumption of adequate microphysics and algorithms, this density profile is the major determinant of explosion for non-rotating progenitors.  This perspective emphasizes the central role of stellar evolution prior to the core Chandrasekhar instability at the terminal phase of a massive star.  However, there exist in these profiles numerous features that will in principle affect the competition between ``$L_{\nu}$" and $\dot{M}$ or between $t_{adv}$ and $t_{heat}$.  Among these are the depths and positions of the various shell interfaces, the central densities at collapse, and the slopes of the density profiles along the smooth segments throughout the inner region interior to $\sim$2.5 M$_{\odot}$. There are in principle numerous relevant parametrizations for these profiles.  What might be the correlation between any one of these quantities and explodability?  This is what we set out to research, though we again emphasize that our conclusions are constrained by whatever limitations our current F{\sc{ornax}} simulations may embody.  We also note that the wide range in general structures, from steep outer profiles for the less massive progenitors to much more shallow outer profiles for the more massive progenitors, is not rigorously monotonic with ZAMS mass, but only crudely so \citep{sukhbold2018}. The mapping between ZAMS mass and terminal density profile may be as subtle and ``chaotic" as the current literature suggests, but the general range of density profiles depicted in Figure \ref{fig:rho-M} may reflect reality.  However, one should keep in mind that the detailed mapping between ZAMS mass and terminal profile is still a work in progress \citep{Chatzopoulos2016,muller2016,muller2017,Muller2019,fields2020,Fields2021}. 
 
\begin{figure*}
    \centering
    \includegraphics[width=0.47\textwidth]{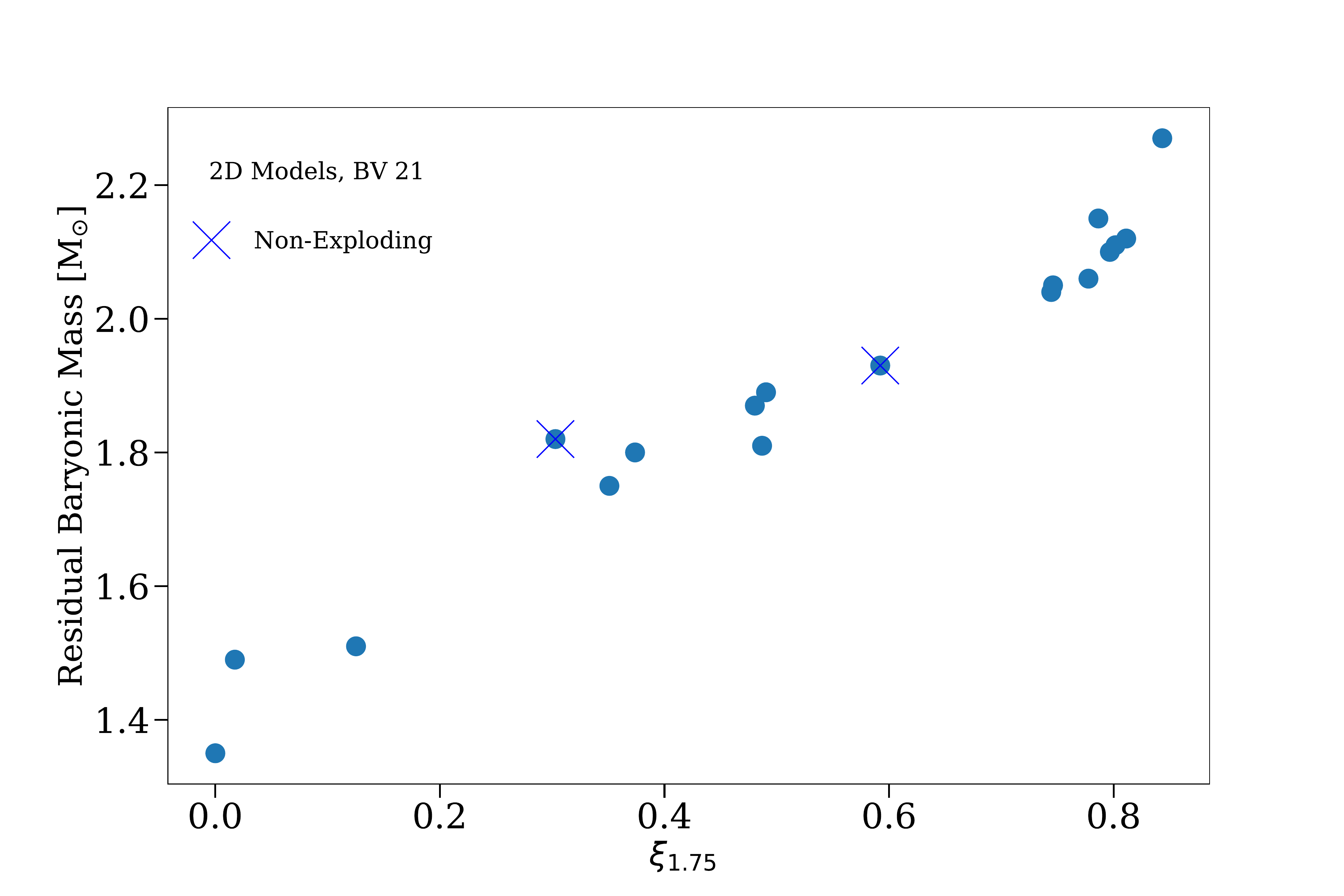}
    \includegraphics[width=0.47\textwidth]{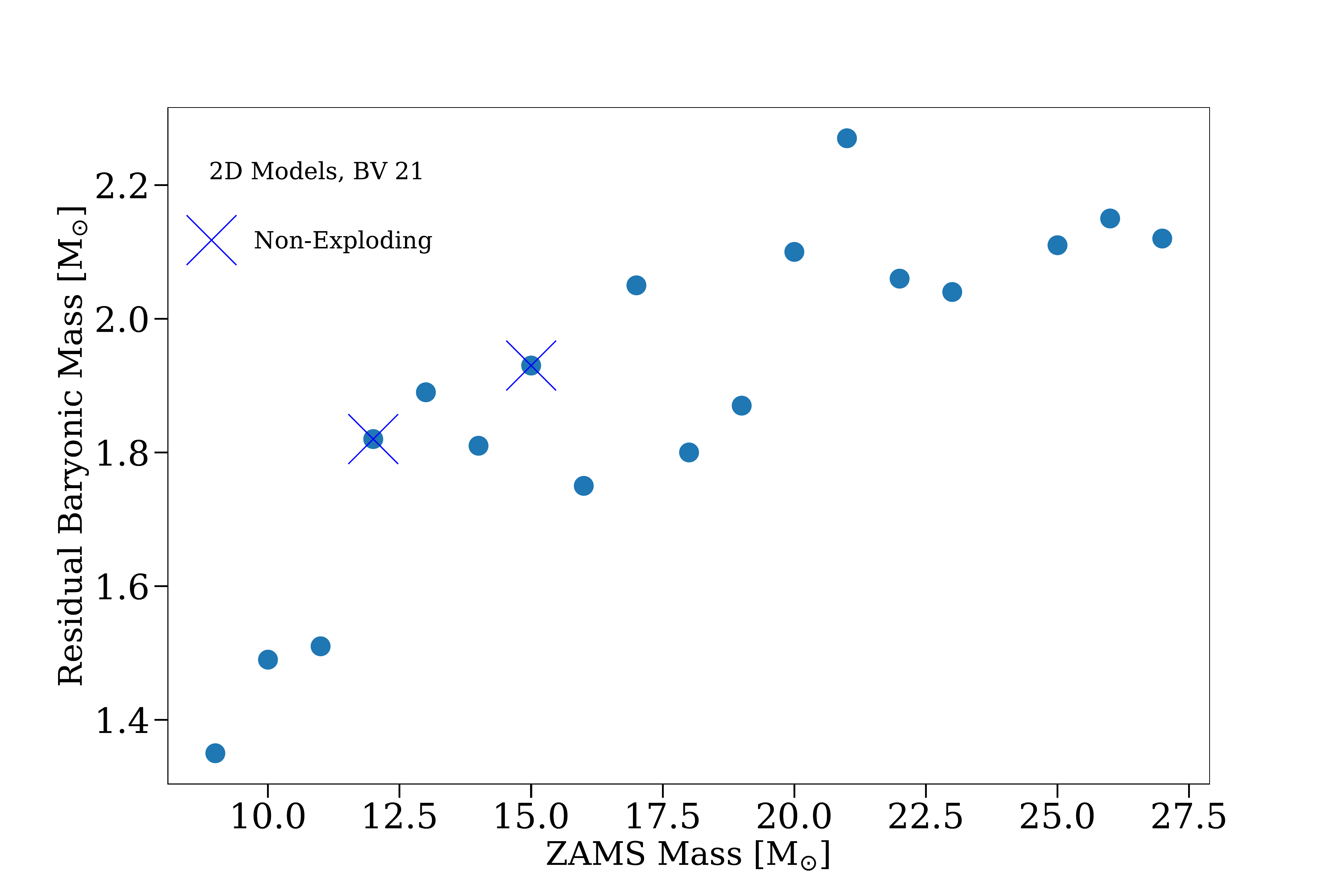}
    \includegraphics[width=0.47\textwidth]{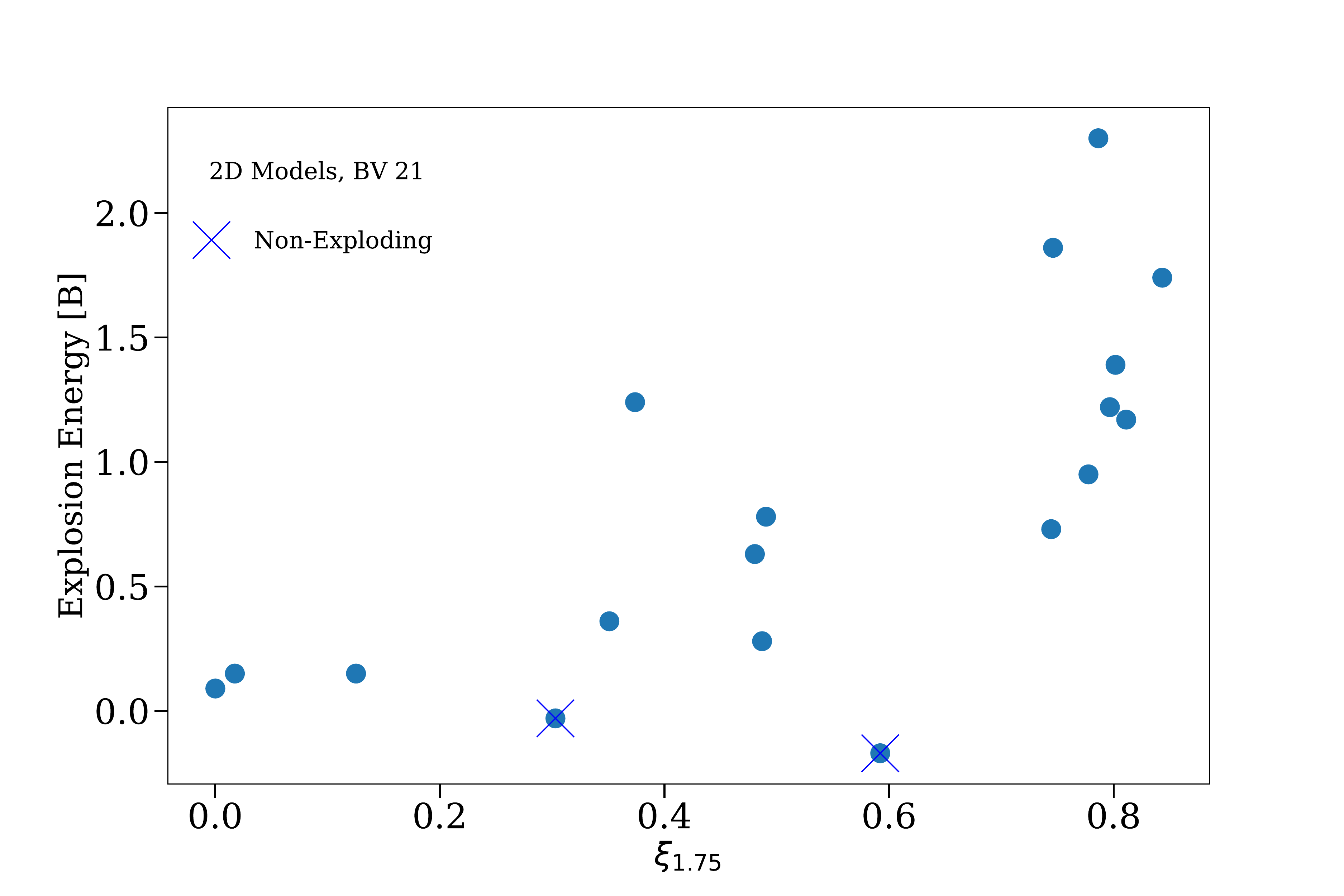}
    \includegraphics[width=0.47\textwidth]{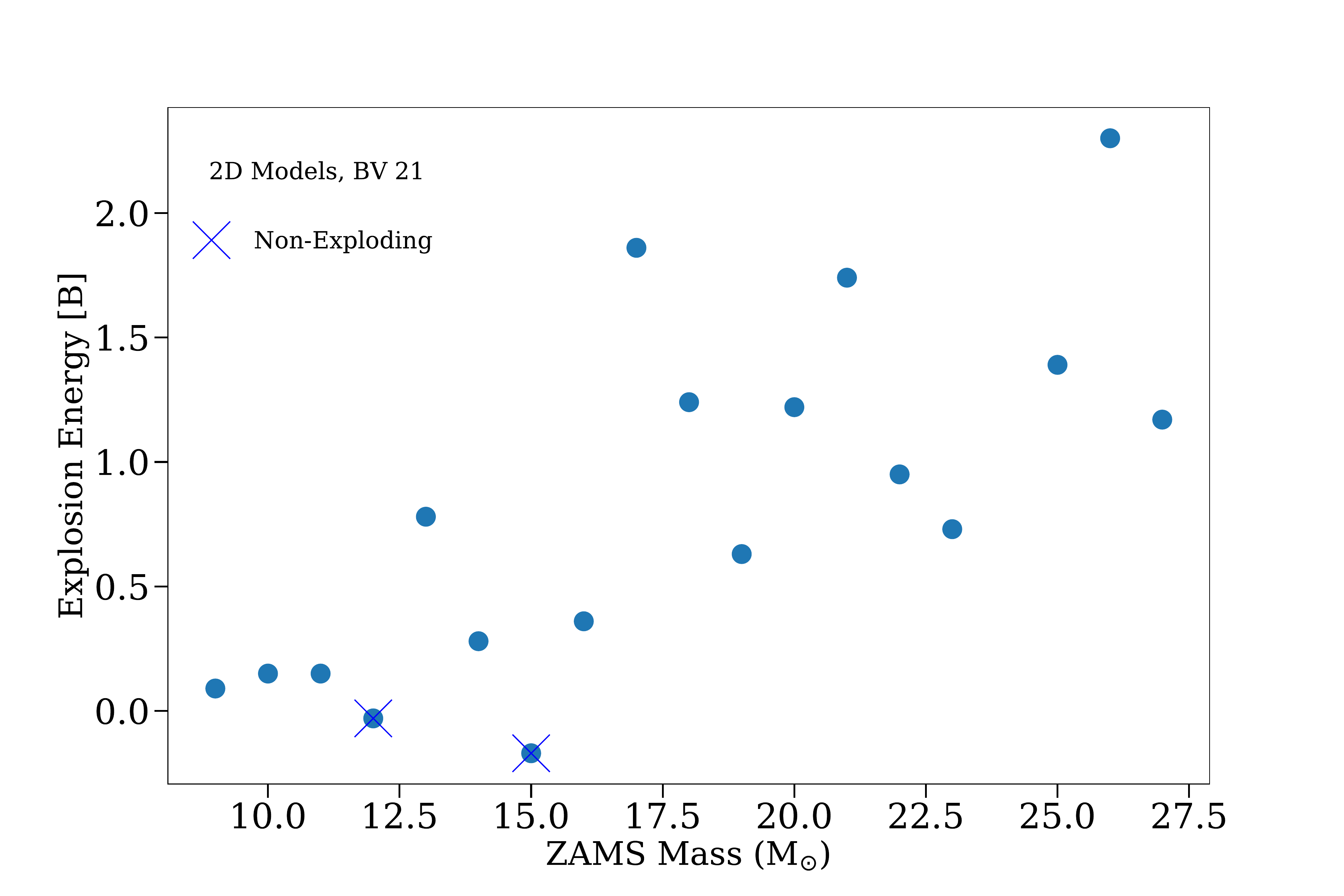}
    \caption{Residual baryonic PNS mass in M$_{\odot}$ ({top}) and explosion energy in Bethes ({bottom}) as a function of progenitor compactness (at 1.75 M$_{\odot}$) ({left}) and progenitor mass ({right}) from the set of 2D late-time simulations published in \protect\cite{2021Natur.589...29B} (``BV 21") and using the \protect\cite{sukhbold2018} progenitors. Models 12- and 15-M$_{\odot}$ do not explode (marked with ``X"s). We identify a strong trend with compactness of PNS mass (in particular) and explosion energy.  There is a corresponding trend with progenitor ZAMS mass, but it is weaker \citep{2021Natur.589...29B,oconnor2013}. See text for a discussion.}
    \label{fig:profile}
\end{figure*}

Given this, we note some of the interesting systematic dependences we do see in the 2D model set between residual PNS baryon mass and asymptotic explosion energy (y-axes) and the compactness and the ZAMS mass (x-axes).  These are portrayed in Figure \ref{fig:profile}.  We see that the final PNS mass is strongly correlated with compactness ($\xi_{1.75}$) and, with a bit more noise, with ZAMS mass, while the explosion energies are less correlated with either.  However, there is a hint that low-mass, low-compactness progenitors are less energetic than high-mass, high-compactness progenitors.  The former trend is to be expected, since the post-bounce mass accretion rates increase reliably with compactness, and the latter is reasonable, since high compactness boosts both the driving accretion neutrino luminosities and the gain region mass upon which a fraction of the emerging neutrinos are absorbed.  

\begin{figure*}
    \centering
    \includegraphics[width=0.6\textwidth]{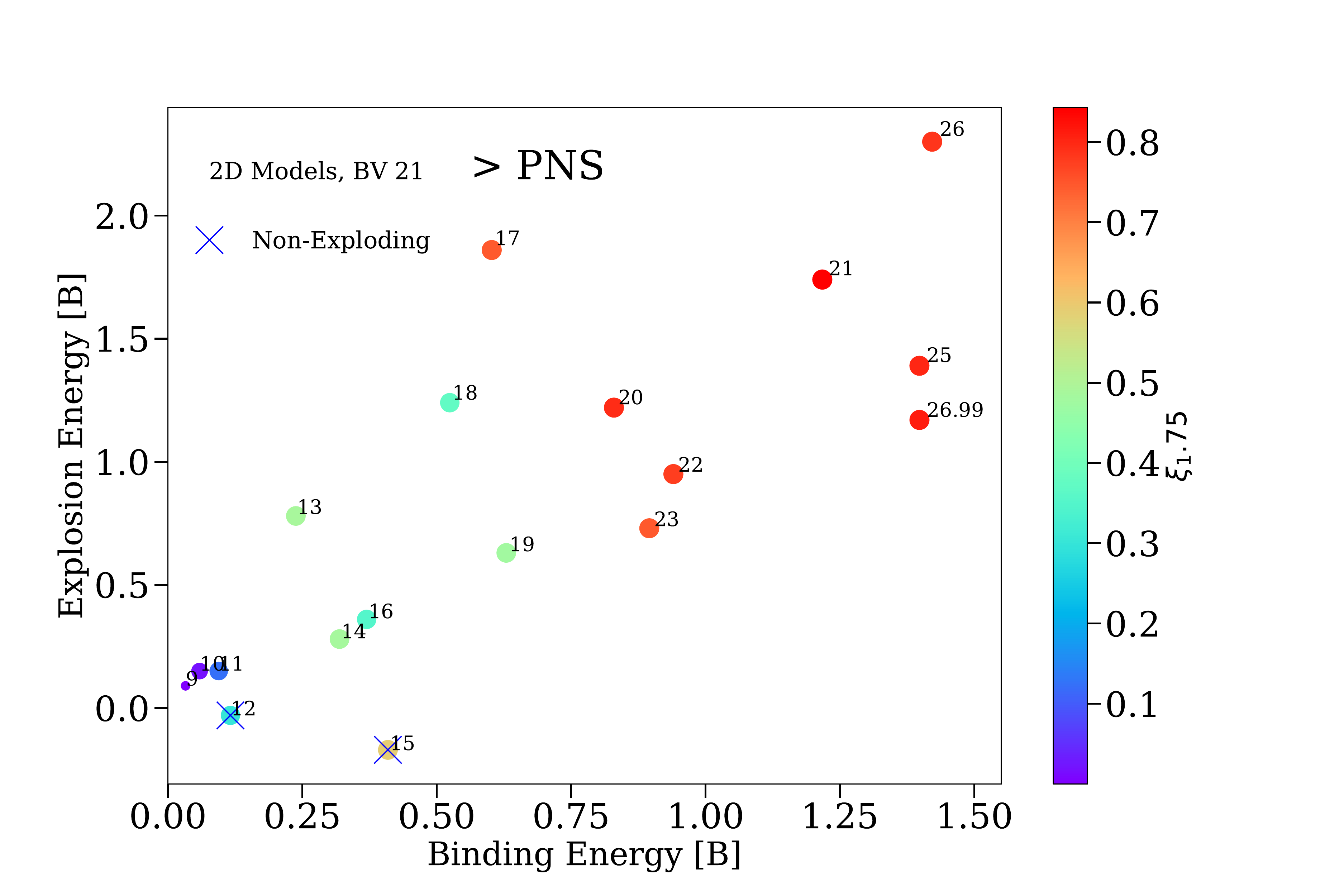}\caption{The explosion energy (in Bethes) versus the initial binding energy of the progenitor exterior to the residual PNS baryon mass (in Bethes) for the 2D models published in \protect\cite{2021Natur.589...29B} (``BV 21"), colored by compactness.  Each model is identified by its ZAMS mass and colored by its compactness. Non-exploding models are indicated with ``X"s.  A clear trend, with a large spread, is in evidence.}
    \label{fig:Eexp-Eb}
\end{figure*}

\begin{figure*}
    \centering
    \includegraphics[width=0.40\textwidth]{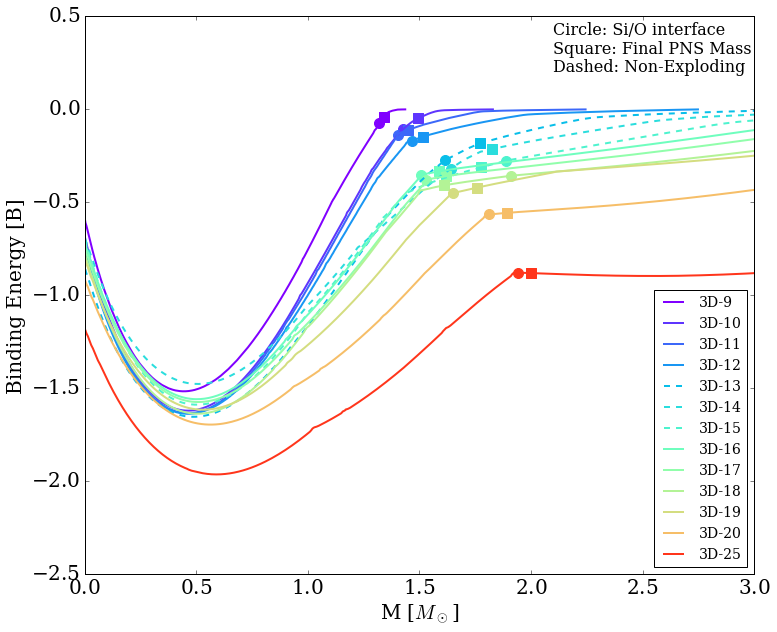}
    \includegraphics[width=0.40\textwidth]{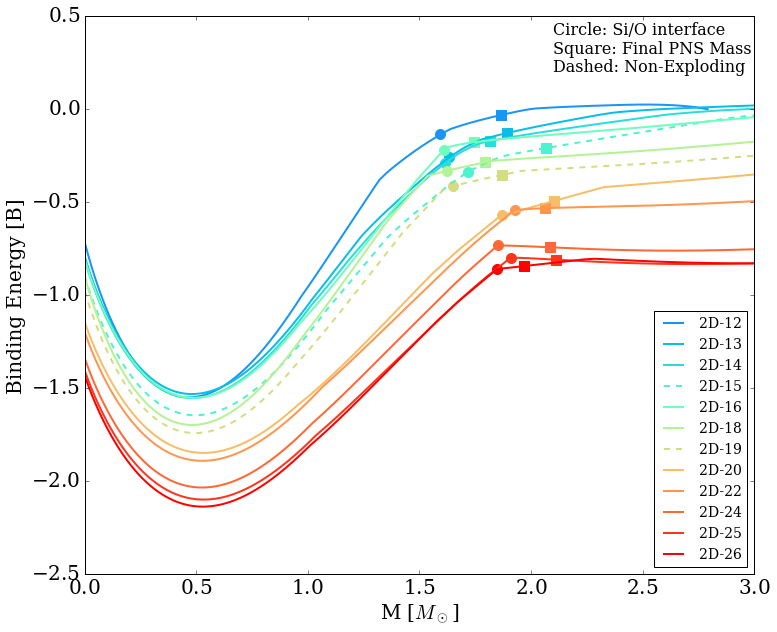}
    \caption{The exterior binding energy of the massive-star progenitors in Bethes (10$^{51}$ ergs) versus interior mass (in M$_{\odot}$). Left: 3D simulations;  Right: 2D simulations. The solid lines are exploding models, while the dashed lines are the non-exploding models. The circular dot on each line shows the position of the Si/O interface, while the square shows the PNS mass at the end of each simulation.}
    \label{fig:eb}
\end{figure*}

Figure \ref{fig:Eexp-Eb} renders the dependence of the explosion energy on the mantle binding energy and suggests that this binding energy does indeed help to set the energy scale of explosion (when models do explode). This figure simultaneously recapitulates the trends with compactness and ZAMS mass in Figure \ref{fig:profile}, emphasizing again the systematics, loose or tight, between these quantities. Figure \ref{fig:Eexp-Eb} is an important possible finding of modern CCSN theory, but should be viewed as provisional
until verified more broadly.  The binding energy versus interior mass for the host of 3D and 2D models of our ``training set" are given in Figure \ref{fig:eb}.  This figure suggests \tianshu{that there is} a general expectation connecting the ``mass cut" \tianshu{(interior to which all materials will be left behind after the explosion)} and the explosion energy in the context of the possible correlations depicted by Figures \ref{fig:profile}, \ref{fig:Eexp-Eb}, and \ref{fig:eb}.

Using these heritage 2D and 3D models, we attempted to find indices of explodability that might be used to map supernova explosion outcome (a binary ``yes/no") with initial density profile (Figure \ref{fig:rho-M}). We explored numerous approaches, but settled on one which we now describe in \S\ref{motivation}. This condition, while not perfect, seems to have predictive value, though there are a few exceptions that need explaining.

\section{Motivation for Our Explosion Criterion}
\label{motivation}

\begin{figure*}
    \centering
    \includegraphics[width=0.40\textwidth]{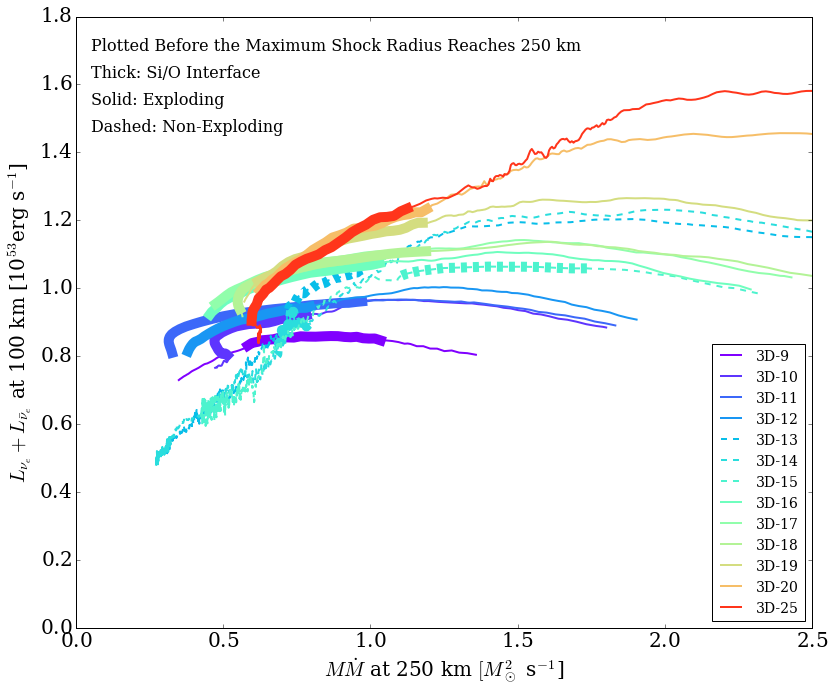}
    \includegraphics[width=0.40\textwidth]{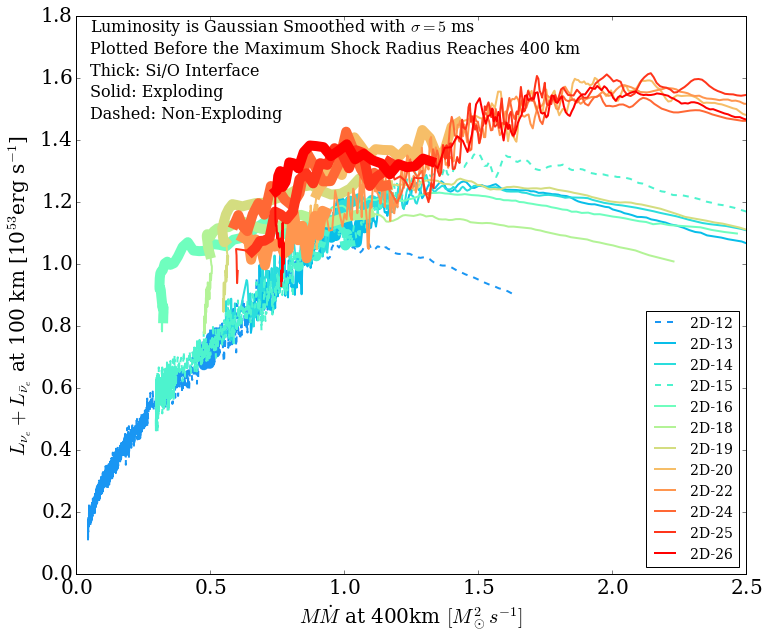}
    \includegraphics[width=0.40\textwidth]{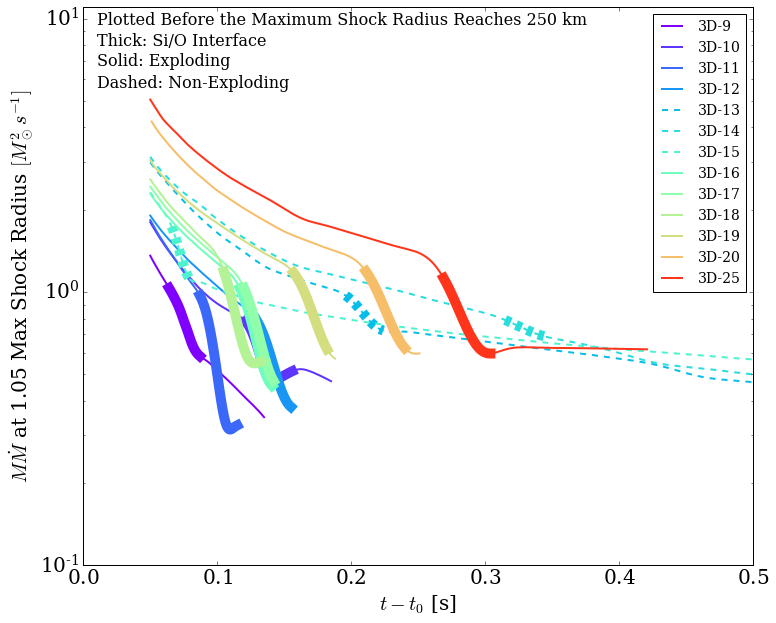}
    \includegraphics[width=0.40\textwidth]{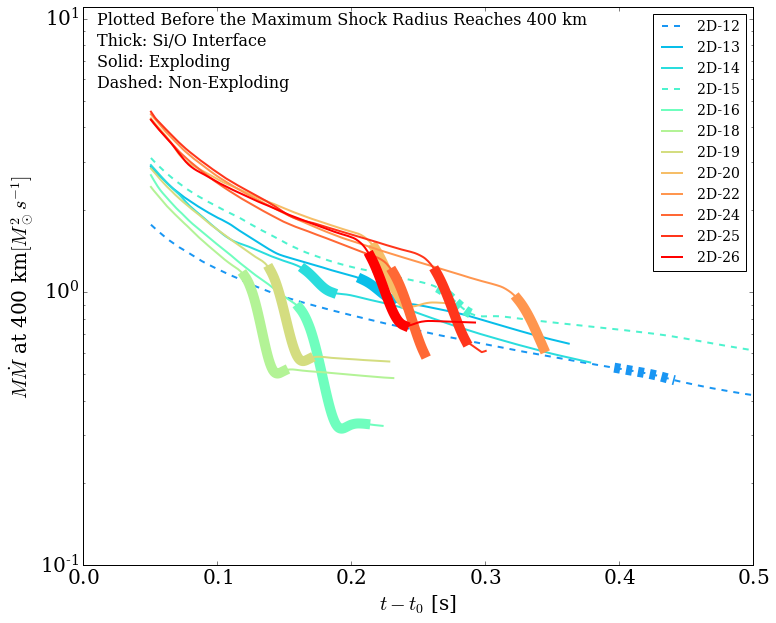}
    \caption{The sum of the electron and anti-electron neutrino luminosities (at 100 km) versus $M\dot{M}$ (top) and $M\dot{M}$ versus t (time after bounce) (bottom).
    The left plots are for the 3D models and the right plots are for the 2D models. The accretion rates ($\dot{M}$) and interior masses ($M$) are measured at 250 km (for 3D) and 400 km (for 2D), i.e., at the radii shown by the dashed horizontal curves in Figure \ref{fig:rshock}. As before, the 3D simulations are from \protect\cite{burrows2020}, while the 2D simulations are from \protect\cite{2021Natur.589...29B}. Solid lines are exploding models, while dashed lines are non-exploding models. The thick portion of each curve marks the density discontinuity (and, hence, the discontinuity in the accretion rate) at the Si/O interface, which is more clearly shown in the bottom panels.  In these plots, we show only the phase before the maximum shock radius reaches 250 km or 400 km so that the accretion rate shown is not influenced by the explosion. From the top panels, we see that the exploding and non-exploding models behave differently. There is a roughly diagonal relation that all simulations (both 2D and 3D) adhere to at later times, and only the exploding models deviate (to the left on the top plots) from that diagonal relation. See text for a discussion.}
    \label{fig:L-mmdot}
\end{figure*}

\begin{figure*}
    \centering
    \includegraphics[width=0.40\textwidth]{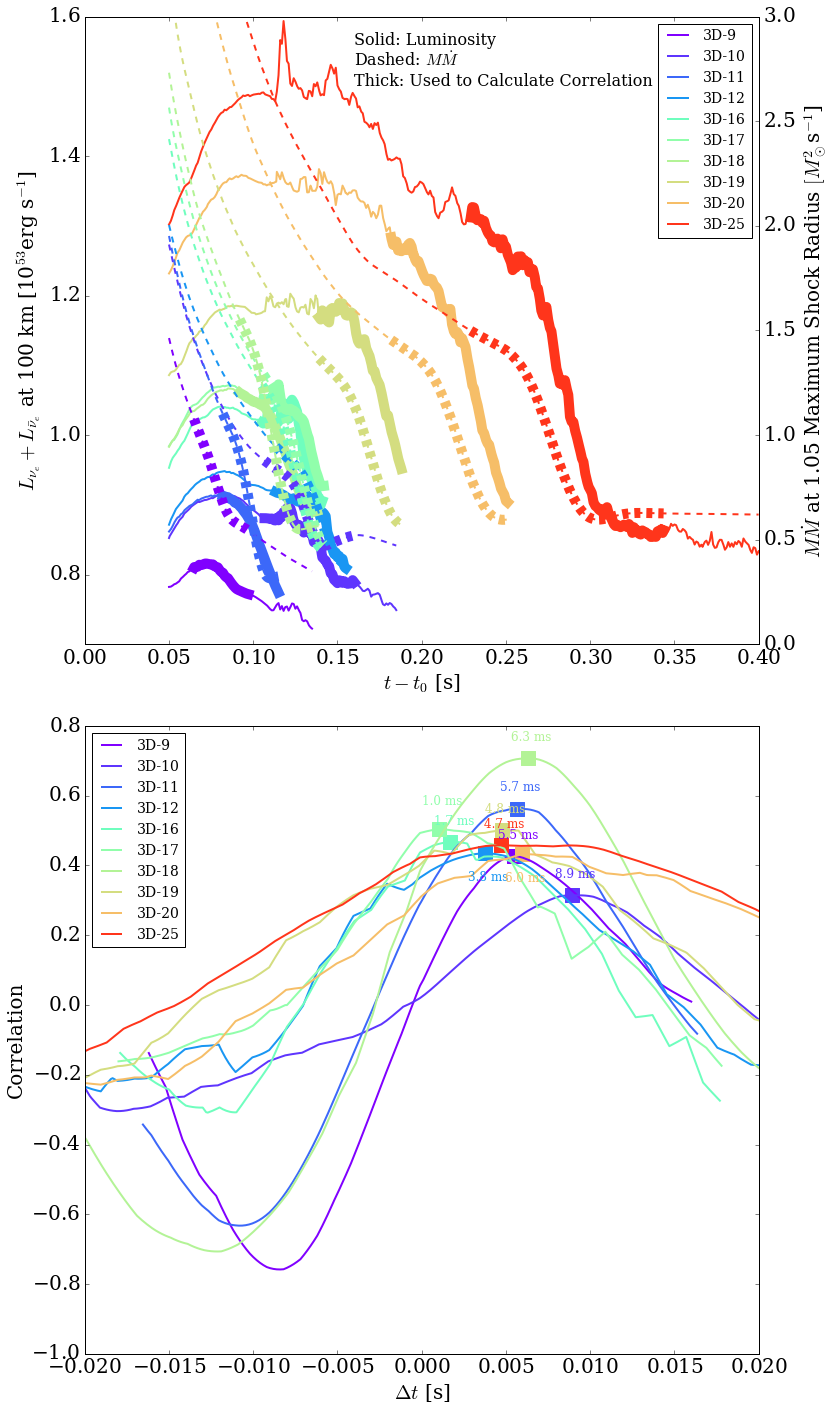}
    \includegraphics[width=0.40\textwidth]{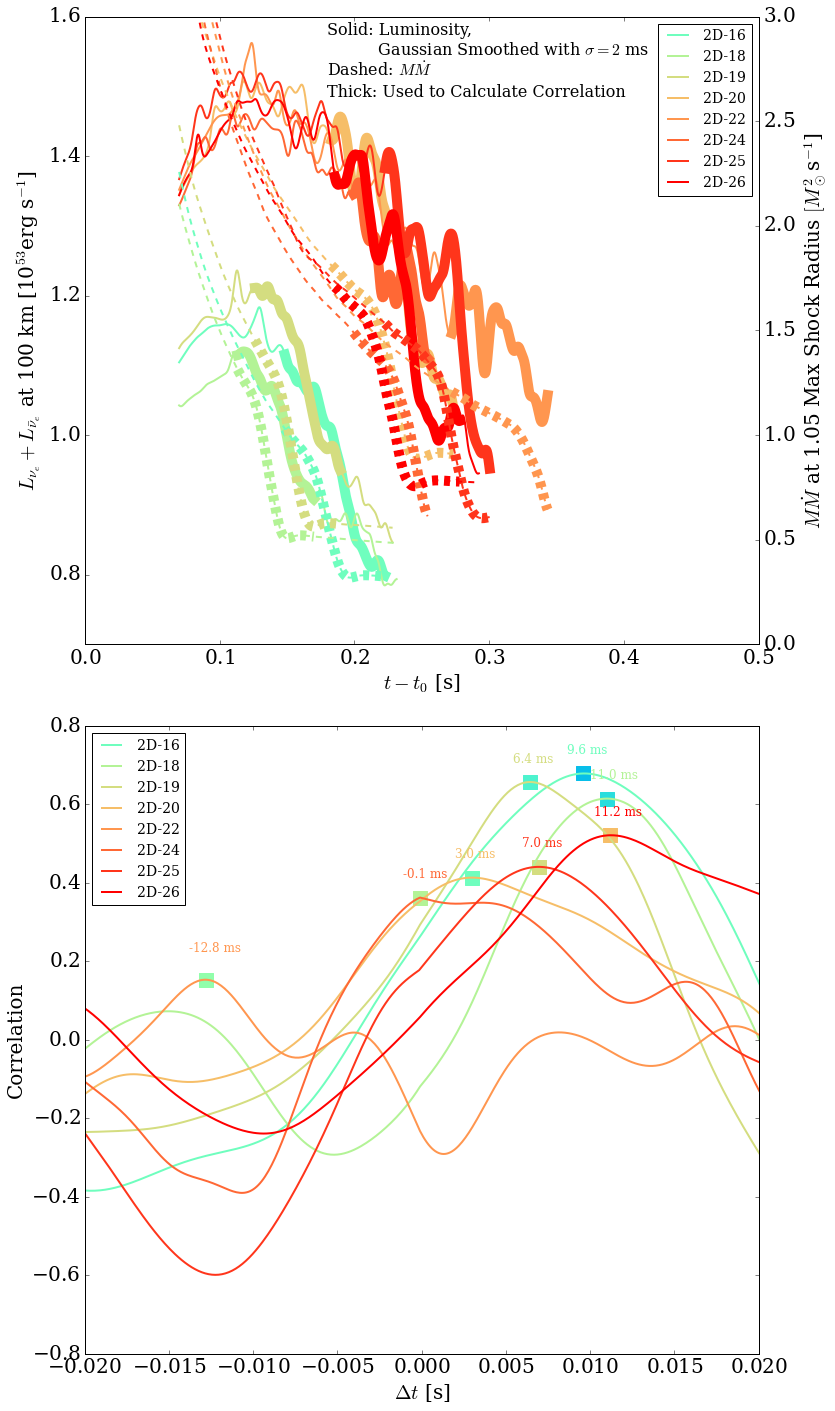}
    \caption{Top panels show the luminosity and accretion rate (times $M$) as a function of time for the 3D (left) and 2D (right) model sets. The solid lines are the sum of the $\nu_e$ and $\bar{\nu}_e$ luminosities (left ordinate), while the dashed lines are $M\dot{M}$ (right ordinate). The bottom curves render the luminosity-$M\dot{M}$ correlation functions. We use the thick parts of the top curves to calculate this correlation. Although the uncertainty of this time delay measurement is largely due to noisy luminosities (particularly in 2D), we still see that the luminosity lags the accretion rate near when the Si/O interface (or any significant interface) is accreted. Such a time delay causes excursions from the diagonal relation in Figure \ref{fig:L-mmdot} that are often germane to the onset of explosion.}
    \label{fig:delay}
\end{figure*}

Motivated by the general concept of the critical condition between luminosity and mass accretion rate, we extracted from our reference suite of 3D and 2D simulations $L_{\nu_e}$, $L_{\bar{\nu}_e}$, and $\dot{M}$ and plotted in Figure \ref{fig:L-mmdot} $L_{\nu_e}$ + $L_{\bar{\nu}_e}$ versus $M\dot{M}$ for these 3D (top, left) and 2D (top, right) models.  It is the $\nu_e$ and $\bar{\nu}_e$ neutrinos that are responsible for the lion's share of the heating in the gain region. Since $\dot{M}$ is a monotonically-decreasing function of time, time is increasing to the left.  Scrutiny of \tianshu{the top panels} in Figure \ref{fig:L-mmdot} reveals that the exploding models jump up and to the left from the general diagonal trend, the latter defined more clearly by the trajectories executed by the non-exploding models.
This evolution is reminiscent of the expected behavior of an exploding model upon exceeding a critical curve.  Interestingly, the thickened portions of the curves trace the phase of the accretion by the shock of the Si/O interface.  This is seen more clearly for each of these models in the lower panels of Figure \ref{fig:L-mmdot}, which depict $M\dot{M}$ versus time after bounce.  The exploding models show a marked declivity, while for those that don't explode this feature is more muted. The corresponding plot versus time for  $L_{\nu_e}$ + $L_{\bar{\nu}_e}$ is given at the top of Figure \ref{fig:delay}.

This difference between exploding and non-exploding trajectories suggests a common role for the accretion through the shock of the density drop of this (or a similar) interface in often igniting a neutrino-driven explosion.  Upon encountering the shock, this rather abrupt drop in density translates into an immediate  drop in ram pressure.  However, the corresponding drop in the accumulation of mass in the deep core is delayed by a time related to $t_{adv}$ and it is in the deeper core that the accretion component of the luminosity is realized.  The bottom of Figure \ref{fig:delay} portrays the correlation function between the summed luminosity and $M\dot{M}$ versus the time delay ($\Delta t$). \tianshu{This correlation function is calculated using $\langle(f(t)-at-b)(g(t+\Delta t)-c(t+\Delta t)-d)\rangle$, where $a$, $b$, $c$, $d$ are the parameters of the best linear fits for the functions $f$ and $g$.}  There is a bit of noise in these curves that can compromise interpretation, but overall the delay times are positive.  
Hence, the decrease in the bounding ram pressure slightly leads the associated decrease in the driving luminosity.
During this interval, the trajectories depicted \tianshu{in the top panels} of Figure \ref{fig:L-mmdot} can jump \tianshu{leftward\footnote{In the top panels, the trajectories proceed from top right to bottom left.}}.  If there is a proximate critical curve, then such a jump can lead to explosion.  Once explosion commences and the matter starts to expand, the cooling component in the gain region subsides, while the heating rate is more stable.  Therefore, the instability can run away. The accretion of the interface and the relative delay in the response of the neutrino luminosity facilitate explosion. Note that in multi-dimensional simulations the radial position of the shock wave is angle-dependent. As a result, the accretion time of an interface is not simultaneous. This complicates its role and this interpretation, but we surmise not by much.

\begin{figure*}
    \centering
    \includegraphics[width=0.8\textwidth]{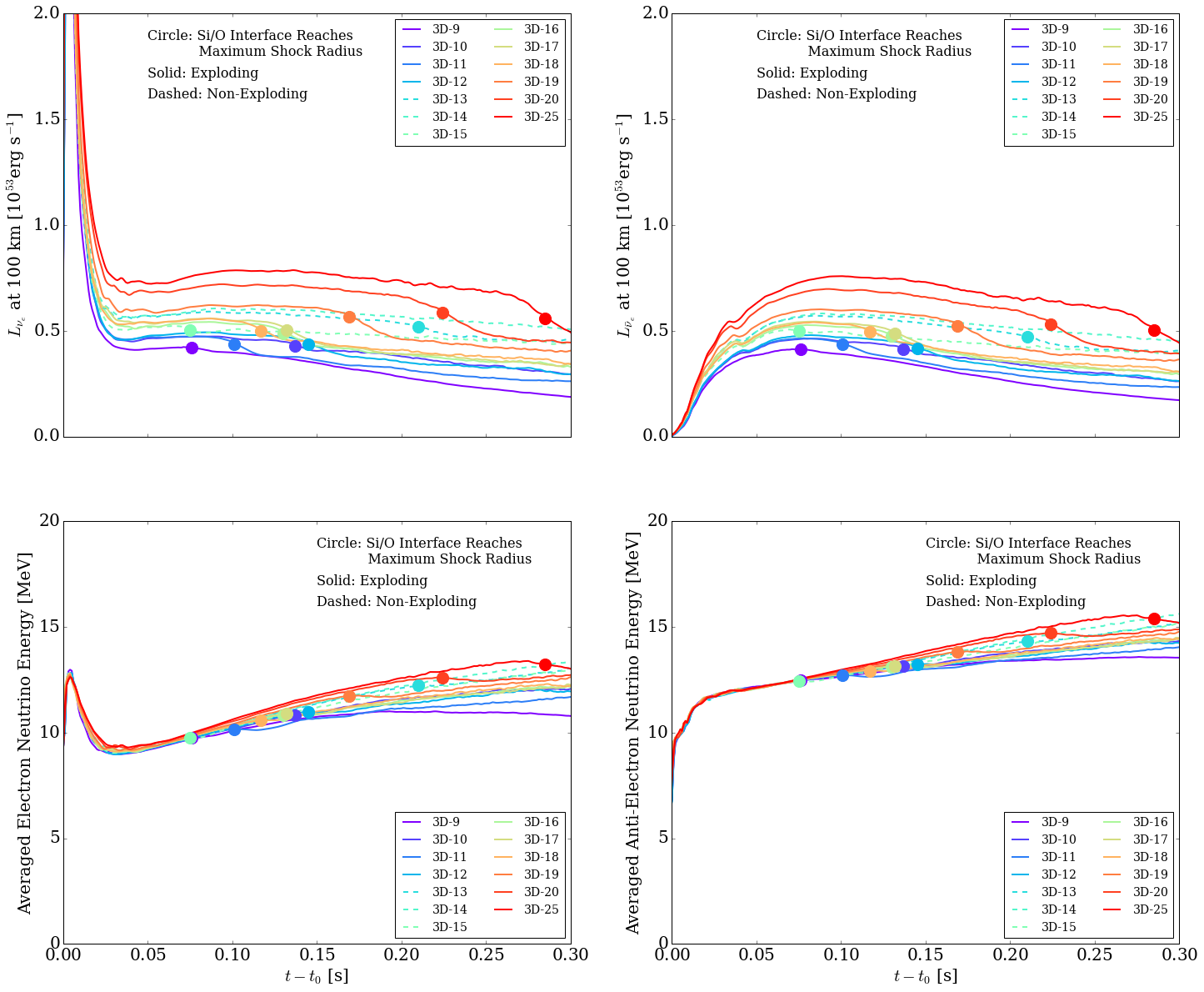}
    \caption{Electron neutrino (left) and anti-electron (right) neutrino luminosity and average neutrino energy versus time after bounce for the reference 3D simulations. The top panels are luminosities measured at 100 km and the bottom panels are the average neutrino energies. The solid lines trace the exploding models, while the dashed lines trace the non-exploding models. The dot on each line marks the time when the Si/O interface is accreted through the shock. We see that the average electron neutrino energy first decreases, and then increases, with time after bounce. For the anti-electron-types, the average energy only increases during the first many hundreds of milliseconds after bounce. Since the cross sections are proportional to the square of the neutrino energy, the neutrino heating rates track these behaviors for a given luminosity. Hence, if the Si/O interface meets the shock too early, the heating rate could be too low to support an explosion. See the text for a discussion.}
    \label{fig:avg_nue}
\end{figure*}

In Figure \ref{fig:avg_nue}, we plot for the ``training" 3D and 2D simulations the luminosity and average neutrino energy evolution after bounce during the first 300 milliseconds for the $\nu_e$ and $\bar{\nu}_e$ neutrinos.  These panels show not only that the plateau luminosity is \tianshu{approximately} monotonic with progenitor mass (and compactness), but that the $\bar{\nu}_e$ luminosity takes almost $\sim$3$-$50 milliseconds to achieve its level. Before this time it is not substantially contributing to neutrino heating in the gain region.  This suggests that, unless $\nu_e$-type neutrinos alone can ignite explosion, if the Si/O (or any other) interface is accreted too early explosion by this route may be inhibited. Figure \ref{fig:avg_nue} also shows that during the crucial early hundreds of milliseconds after bounce the average neutrino energies are increasing. Since the neutrino-matter absorption cross sections increase roughly as the square of the neutrino energy, due to this factor alone the mantle power deposition grows with time and the mantle is getting more explodable. This trend and the early lethargy in the growth of $L_{\bar{\nu}_e}$ together delay explosion for many models.
It is only, we suggest, for those models with very steep outer mass density profiles
and the lowest compactness ratios that these considerations might be less important.

Importantly, these considerations are made in the context of our F{\sc{ornax}} simulations.  Though we have incorporated a complicated suite of microphysical processes (including inelastic scattering on electrons and nucleons); weak magnetism and many-body corrections; velocity-dependent and multi-group transport; and approximate general relativity \citep{skinner2019,vartanyan2019,burrows2020}, no doubt there are limitations to our calculations, as there are limitations in the initial models.  In addition, the outcome of the simulations will depend upon spatial and energy-group resolution \citep{nagakura2019b}. The degree to which these limitations undermine our findings is unknown, but we believe our simulations and the results that have emerged from them are competitive and informative.

\section{Test Explodability Condition}
\label{test}

Given the earlier discussions, we now motivate a possible explodability condition; it  involves the evolution of the ram pressure at the shock and the effect of the accretion of the Si/O (or other) interface and its associated density drop.  Exceptions to the direct predictability of this condition may be a very low compactness parameter (leading to early explosion) or the very early accretion of an interfacial density drop (forestalling explosion). It may also be that the region in the space in which we suggest one look to determine explodability possesses a fuzzier band than we suggest separating the exploding models from the non-exploding models. This condition is derived from the initial density profile of the progenitor alone and does not require a detailed simulation capability.  However, it is calibrated on our more detailed F{\sc{ornax}} results (\S\ref{explore};\S\ref{motivation}). 

We assume that two of the main conditions for explosion are (1) a strong density discontinuity at the Si/O interface (as long as it is not accreted through the stalled shock too early after bounce) and/or (2) a steep density profile (e.g., that for the 3D 9$M_\odot$ model). We suggest that progenitors that meet either of these two conditions will probably explode. 

In an approximate sense, the density profiles become progressively less steep and the density jumps at the Si/O interfaces become more pronounced when transitioning from low-mass progenitors to high-mass progenitors. Progenitors that have both a steep density profile and a strong Si/O discontinuity (like the 3D-11 model) will explode very easily and usually quite early. Progenitors with a steep density profile, but without a strong discontinuity (or the discontinuity falls in too early), also explode, but there won't be a sudden shock jolt caused by the discontinuity. Progenitors without a steep density profile, but with a strong Si/O discontinuity, will explode when the interface reaches the shock, even if the shock is already shrinking in average radius (as with the 3D-25). But if the interface falls into the shock too early, the luminosity and the averaged neutrino energy are still growing (particularly the $\bar{\nu}_e$ luminosity, so the neutrino heating rate may not yet be strong enough for an explosion (witness the case of 3D-15). Progenitors that have neither a steep density profile nor a strong discontinuity seem hard to explode. Examples of such progenitors are the 3D-13 and 3D-14.

A measure of the strength of the density discontinuity is the ram pressure jump when the discontinuity reaches the shock. If the pressure jump is large enough, the shock is kicked into explosion. A simple estimate of the ram pressure at the shock radius can be obtained using the following method: We assume that the shock radius is at $R_s=200$ km and that the infall velocity is given by the free-fall velocity $v(M)=\sqrt{\frac{2GM}{R_s}-\frac{2GM}{r(M)}}$. We derive an approximate fitted relation to estimate the density \tianshu{just exterior to} the shock $\rho_s(M)=\frac{r(M)}{R_s}\rho(M)$, where $r(M)$ and $\rho(M)$ are the initial radius and density of a mass shell. Then, the ram pressure is given by $P_{\rm ram} =\rho_s(M) v(M)^2$.  The infall time is \tianshu{estimated} by a fitted fractional free-fall time $t(M) =\sqrt{\frac{\pi}{4G\bar{\rho}(M)}}$ \tianshu{\citep{2016MNRAS.460..742M}}, where $\bar{\rho}(M) =\frac{3M}{4\pi r(M)^3}$ is the average density interior to a mass coordinate. The fractional change in ram pressure is then calculated by $\frac{P_{\rm ram}(t)-P_{\rm ram}(t+\Delta t)}{P_{\rm ram}(t)}$, where $\Delta t$ is estimated as $\frac{7R_s}{v(M)}$. The factor of 7 is motivated by the shock jump for a $\gamma = 4/3$ gas.

There is another expression we can use to estimate the ram pressure: $P_{\rm ram}=\frac{v(M)}{4\pi R_s^2}\frac{dM}{dt}$. Although the ram pressure prescriptions may differ, the fractional change across the interface is almost the same, since we are looking at the fractional change in a very short time period $\Delta t<10$ ms, and all slowly changing quantities (such as $v$) do not by such changes influence the result.

\begin{figure*}
    \centering
    \includegraphics[width=0.40\textwidth]{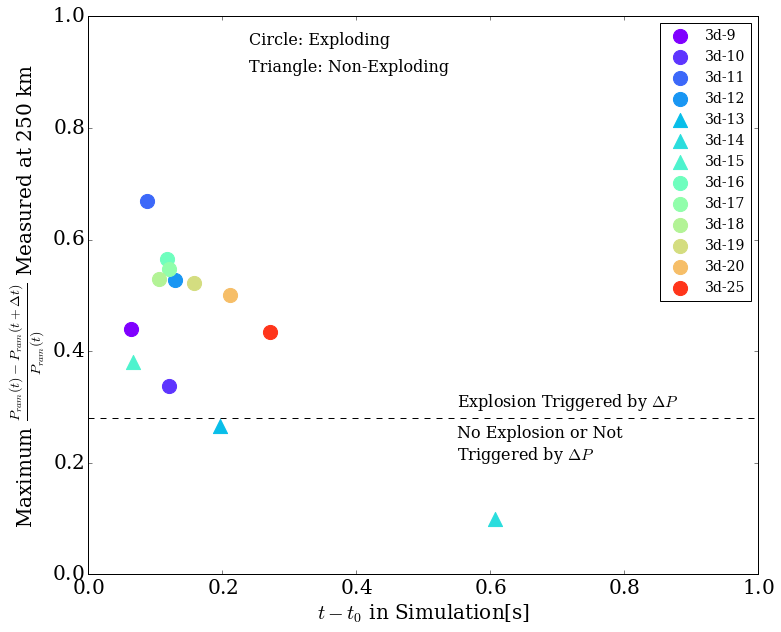}
    \includegraphics[width=0.40\textwidth]{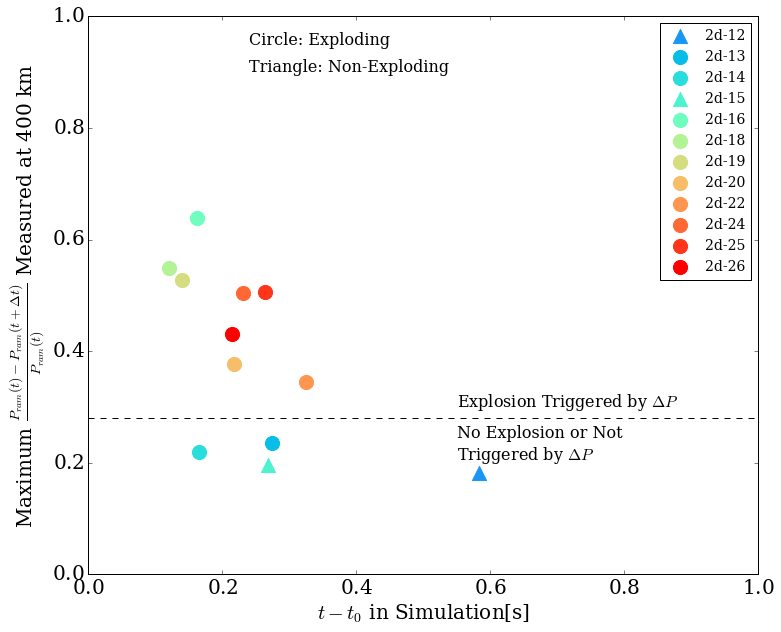}
    \caption{Maximum fractional ram pressure change over a time shift $\Delta t$ measured for the 3D (left) and 2D (right) simulations. The circular dots identify exploding models, while the triangles indicate the non-exploding models. $\Delta t$ is taken to be equal to $\frac{7R_{s,max}}{v}$, where $R_{s,max}$ is the maximum shock radius and $v$ is the infall velocity measured at 250 km (for 3D) and 400 km (for 2D). For this plot, we ignore the pressure change caused by the ram pressure change at bounce. The maximum fractional change generally occurs at the Si/O interface, and the amplitude of this fractional change shows how strongly the discontinuity influences the explosion. From this plot, we can see that the ram pressure changes in non-exploding models are smaller, while for most of the exploding ones they are larger. {We draw a horizontal dashed line at 0.28 to distinguish the strong-discontinuity and weak-discontinuity clusters which can be more clearly seen in Figure \ref{fig:pjump-ini}. Most exploding models satisfy this criterion, with three exceptions (3D-15, 2D-13 and 2D-14). However, when we look at the shock evolution of these three exceptions in Figure \ref{fig:rshock}, we find that the shocks of 2D-13 and 2D-14 expand much more slowly compared to the other exploding models at the time the Si/O interface falls in.} This indicates that either they are marginally explosive or that the explosion is not directly triggered by the ram pressure jump. {The radius of the shock of model 3D-15 expands earlier and faster compared to other non-exploding models, which indicates that it might be marginally explosive or that the interface falls in too early to have enough neutrino heating rate for an explosion.} See text for a discussion.}
    \label{fig:pjump-sim}
\end{figure*}

Figure \ref{fig:pjump-sim} shows the calculation of this explodability condition for the 3D (left) and 2D (right) ``training" models using the actual numbers of the detailed simulations.  Max($\frac{\Delta P_{\rm ram}}{P_{\rm ram}}$) versus $t-t_0$ is plotted for each model. The straight dashed lines are those that might be suggested to separate models that explode (upper) from those that don't (lower) and the circles and triangles depict actual exploding and non-exploding models, respectively.  The condition seems not bad, but not perfect, predicting explodability better than non-explodability (particularly for the 2D models) \footnote{We include $t-t_0$ because when calculating the ram pressure jump we search over the entire progenitor profile. Thus, all interfaces (not only the Si/O) will be considered, and some interfaces can be significantly more pronounced than the Si/O interface. However, those interfaces occur either too early ($t-t_0 < 0$) or too late ($t-t_0 > 1.5\,\, {\rm seconds}$), which makes them less useful for triggering explosion. Those that are accreted too early don't ignite explosion because the driving luminosities have not yet built up. Those accreted too late may not lead to explosion because the driving luminosities have already begun to abate.}.

\begin{figure*}
    \centering
    \includegraphics[width=0.40\textwidth]{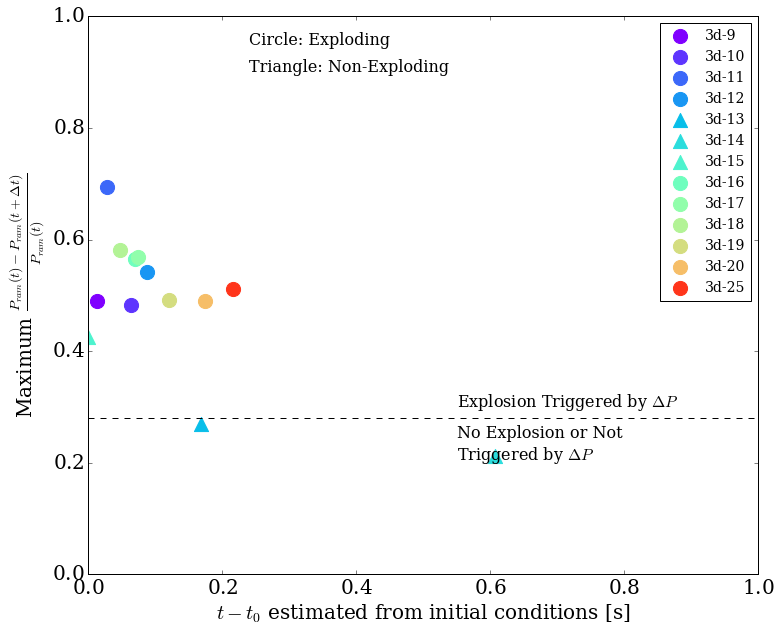}
    \includegraphics[width=0.40\textwidth]{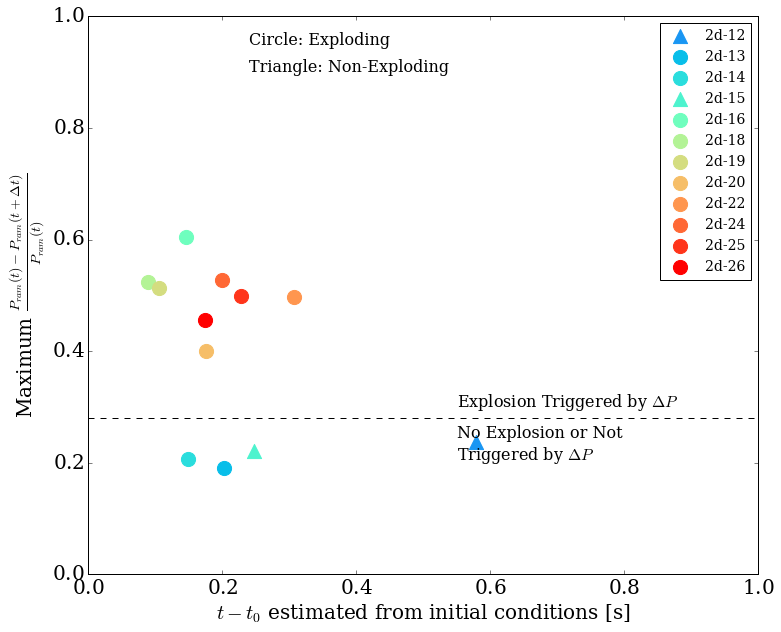}
    \caption{The same as Figure \ref{fig:pjump-sim}, but estimated from the initial progenitor models themselves without reference to the simulations. Circular dots are exploding models, while the triangles are the non-exploding models. $\Delta t$ is given by $\frac{7R_s}{v(M)}$, where $R_s=200$ km. The factor of 7 comes simply from the density jump at a shock for a $\gamma = 4/3$ gas. The infall velocity is estimated using the free-fall velocity $v(M)=\sqrt{\frac{2GM}{R_s}-\frac{2GM}{r(M)}}$. The infall time is calculated using a fraction of this free-fall time fit to the simulations: $t(M)=\sqrt{\frac{\pi}{4G\bar{\rho}(M)}}$, where $\bar{\rho}(M)=\frac{3M}{4\pi r(M)^3}$ is the average density interior to a certain mass coordinate. The bounce time is estimated with the following approximate fitted function: $t_0=0.218\left(\frac{\rho_c}{10^{10}\text{g cm}^{-3}}\right)^{-0.354}$ s. Comparing these plots to Figure \ref{fig:pjump-sim}, we see that this approach predicts explodability quite well. The general trend persists: exploding models tend to have stronger maximum ram pressure jumps, while the non-exploding ones have weaker maximum ram pressure jumps.
    }
    \label{fig:pjump-ini}
\end{figure*}

However, in Figure \ref{fig:pjump-ini} we make the same plot, but use the analytic approximations described above for the same quantities.  Figures \ref{fig:pjump-sim}
and \ref{fig:pjump-ini} look much the same.  Given this, we then ask the question.  If we now calculate using F{\sc{ornax}} with its full physics capabilities 100 2D models and then compare the outcomes (exploding or not) using the analytic approach to ``$\frac{\Delta P_{\rm ram}}{P_{\rm ram}}$" versus time after bounce employing only the initial model mass density profiles how does this explosion condition fare?  

We recapitulate and summarize here our simple prescription for determining the explodability of a model given only the initial density profile:

{
\begin{enumerate}
\item Assume that the shock radius is $R_s=200$ km \tianshu{which is a reasonable estimate for most progenitors}. This assumed radius is too large for interfaces that fall in very early, so we ignore matter interior to the Si/O interface\footnote{The Si/O interface in this paper is defined as the density discontinuity closest to the inner boundary of the oxygen shell in which the oxygen abundance is above 15\%. Sometimes the Si core size given by this definition is underestimated, but this definition does not compromise the accuracy of our criterion.}. More specifically, we ignore data with mass coordinate $M<M_{\rm Si}-0.05M_\odot$.
\item Infall velocity and infall time are given by the free-fall velocity and the fitted fractional free-fall time: $v(M)=\sqrt{\frac{2GM}{R_s}-\frac{2GM}{r(M)}}$ and $t(M)=\sqrt{\frac{\pi}{4G\bar{\rho}(M)}}$, where $r(M)$ is the initial radius of the mass shell and $\bar{\rho}(M)=\frac{3M}{4\pi r(M)^3}$ is the average density interior to it.
\item The density \tianshu{just exterior to} the shock is given by an approximate fitted relation $\rho_s(M)=\frac{r(M)}{R_s}\rho(M)$, where $\rho(M)$ is the initial density of the mass shell. \tianshu{This initial density is extracted when the core collapse speed exceeds a maximum of 900 km s$^{-1}$ \citep{sukhbold2018}.}
\item The ram pressure is assumed given by $P_{\rm ram}=\rho_s(M)v(M)^2$.
\item The time delay is estimated as $\Delta t=\frac{7R_s}{v(M)}$. 
\item The fractional change in ram pressure is then calculated by $\frac{\Delta P_{\rm ram}}{P_{\rm ram}}=\frac{P_{\rm ram}(t)-P_{\rm ram}(t+\Delta t)}{P_{\rm ram}(t)}$. (The infall time $t$ is a monotonically increasing function of mass coordinate $M$, so we can use its inverse function to replace all $M$ dependencies with $t$ dependencies.)
\item Based on our simulation results, we make the following prediction: the progenitor will explode if Max$(\frac{\Delta P_{\rm ram}}{P_{\rm ram}})>0.28$, and vise versa.
\item The bounce time is estimated using a fitted function $t_0=0.218\left(\frac{\rho_c}{10^{10}\text{g cm}^{-3}}\right)^{-0.354}$ s, where $\rho_c$ is the initial central mass density. This is not a very good fit, but it gives us a sense on whether an interface reaches the shock too early or too late.
\end{enumerate}
}

\section{The 2D Model Suite}
\label{suite}
\tianshu{To build this model suite, we employ the SFHo nuclear equation of state \citep{2013ApJ...774...17S} and the same simulation code, F{\sc{ornax}}, we have employed in previous recent investigations. The numerical and physical details incorporated into F{\sc{ornax}} can be found in previous papers \citep{vartanyan2018b,burrows_2019,skinner2019,burrows2020,2021Natur.589...29B}}

\begin{figure*}
\centering
    \includegraphics[width=0.55\textwidth]{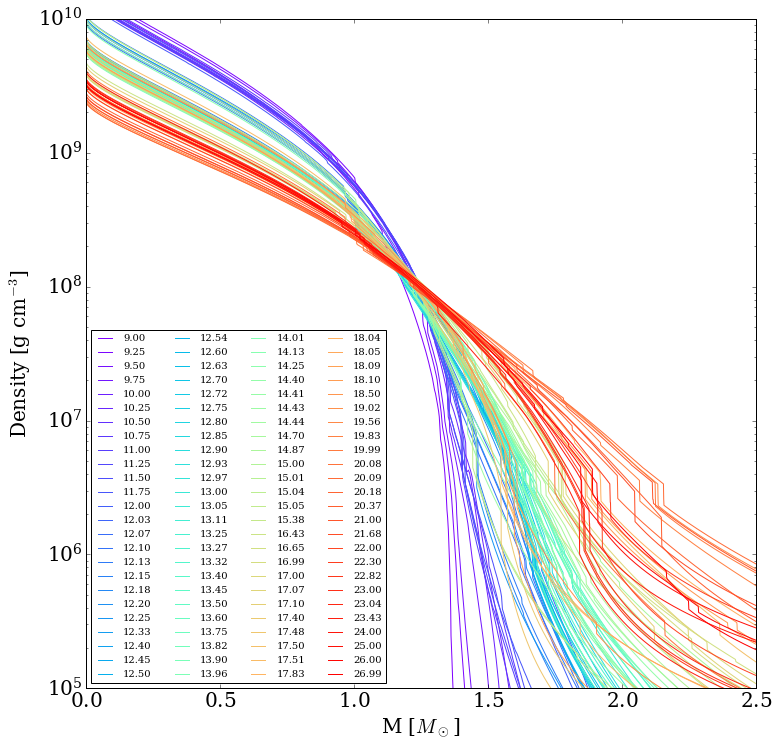}
    \caption{The mass density (in gm cm$^{-3}$) versus interior mass (in M$_{\odot}$) for the 100 progenitors explored in our large 2D study. The models were selected to span the distribution in density space, including both steep and shallow profiles as well strong and weak compositional interfaces.}
\label{fig:density}
\end{figure*}

\begin{figure*}
    \centering
    \includegraphics[width=0.55\textwidth]{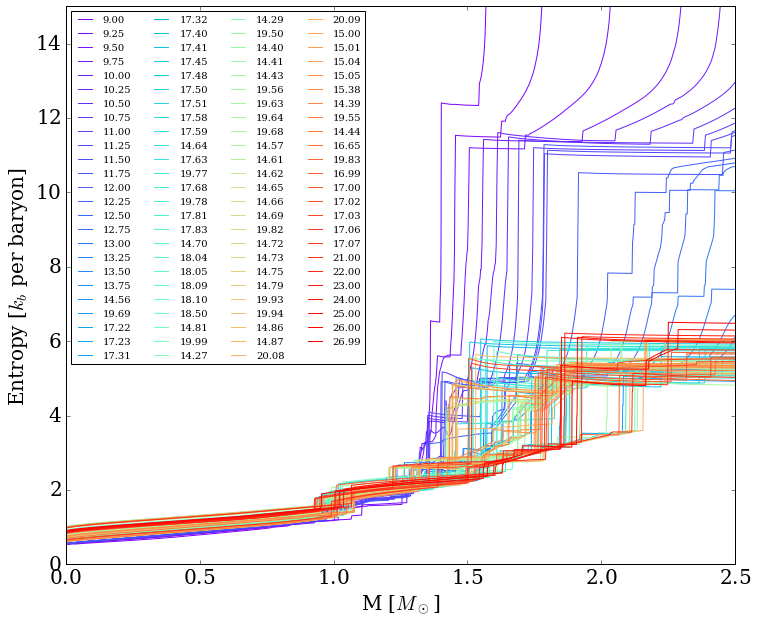}
    \caption{The entropy profiles that correspond to the models depicted in Figure \ref{fig:density}.}
    \label{fig:entropy}
\end{figure*}

The density and specific entropy profiles of the 100 initial progenitor models taken from \citep{swbj16} and \citet{sukhbold2018} are given in Figures \ref{fig:density} and \ref{fig:entropy}. This collection of models clearly spans a wide range of model structures, encompassing 9 to 27 M$_{\odot}$. However, we do not in this study go beyond 27 M$_{\odot}$, leaving that range and (perhaps) black-hole formation to another study.

\begin{figure*}
    \centering
    \includegraphics[width=0.55\textwidth]{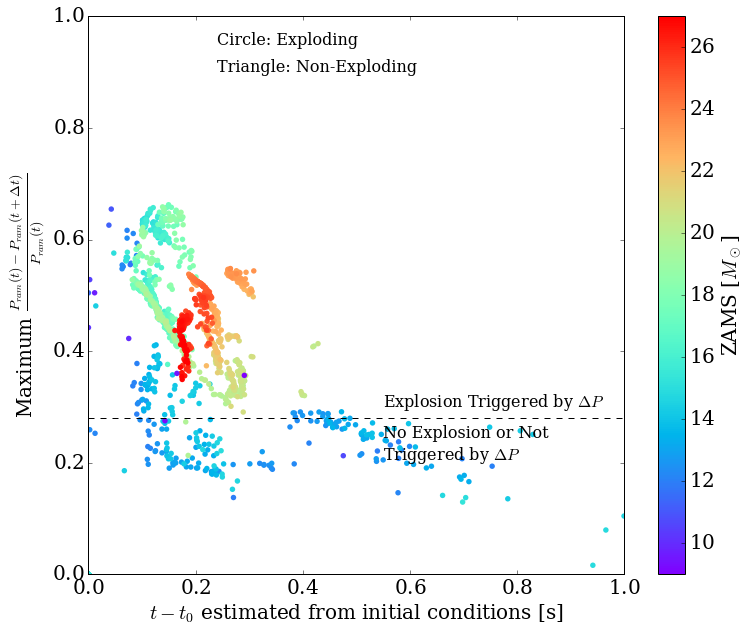}
    \caption{Maximum fractional ram pressure change given a time shift $\Delta t$,  estimated from all the progenitor models we downloaded as a prelude to the large 2D explosion study. Progenitors with $M<12M_\odot$ are from \protect\cite{swbj16}, while progenitors above 12 M$_\odot$ are from \protect\cite{sukhbold2018}. We see that most progenitors with $M>$ 16 M$_\odot$ are above the approximate explosion line, while many low-mass progenitors are below it. However, we would expect that some low-mass progenitors below the line will still explode because of their steep density profiles.}
    \label{fig:pjump-all}
\end{figure*}

Figure \ref{fig:pjump-all} depicts the a priori expectations for the complete set of models in \citet{sukhbold2018}; we use only a fraction of those for our 100-model study.
The color codes for the ZAMS mass, indicating where in progenitor-mass space we might expect duds.

The 100 2D models were simulated out to about 1 seconds after bounce or until explosion, with outer radii from 30,000 to 70,000 kilometers (km).  The default number of radial zones  was 1024 and of angular ($\theta$) zones was 128.  However, we redid numerous models using 256 angular zones.  This was guided by the results in \citet{nagakura2019b}, who suggested models were more explodable at higher resolution.  For a subset of these models this was indeed what we found.  In fact, including the higher resolution cohort
increased the fidelity of the ``Max$\frac{\Delta P}{P}$" test.  

\begin{figure*}
    \centering
    \includegraphics[width=0.55\textwidth]{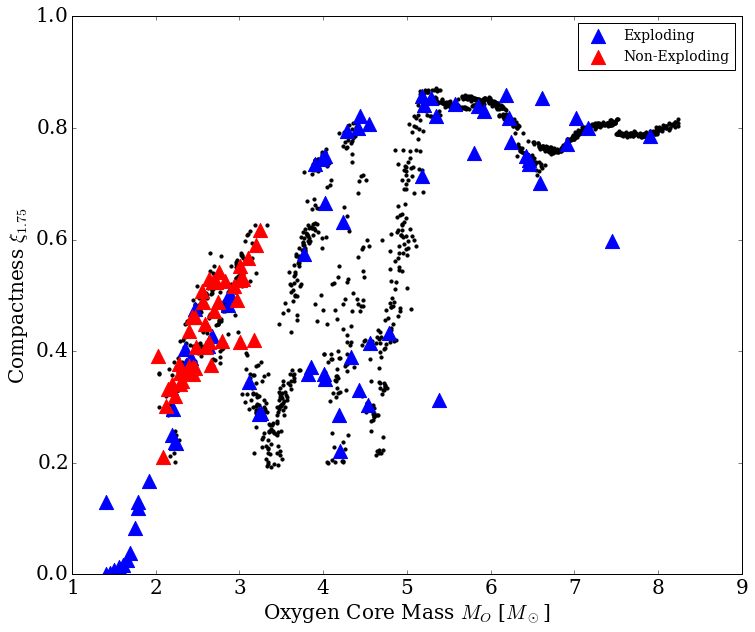}
    \caption{Compactness at 1.75 M$_{\odot}$ versus oxygen core mass. The black dots are a comprehensive set of progenitor models found in the \citet{swbj16} and \citet{sukhbold2018}. Blue and red triangles are the 100 2D simulations in this work. Blue triangles are the exploding models, while red triangles are the non-exploding models. From this figure we see that most of the non-exploding models reside along a specific progenitor branch. We note that in \citet{sukhbold2018} the carbon-oxygen core mass is proportional to the ZAMS mass. This plot is therefore similar to their Figure 15, but contradicts its conclusions concerning which models explode.}
    \label{fig:he-compactness}
\end{figure*}

The results for our 100-model cohort using only initial profile data, but determining explodability using the detailed 2D simulations (``circles" or ``triangles"), is given in Figure \ref{fig:pjump-100}.  This plot is to be compared with Figures \ref{fig:pjump-ini} and \ref{fig:pjump-all}.  The number of false positives is five
in $\sim$40, with six false negatives.  However, the percentage success rate is high.  Moreover, the false positives are all close to the putative demarcation line, suggesting a slightly more nuanced definition may be called for.  We note that the preponderance of the failed models are in the low to intermediate ZAMS mass range and that all the more massive models in the upper progenitor mass range of this study explode.  This is in stark contrast with the expectation of the naive compactness condition. Figure \ref{fig:he-compactness} makes this point more succinctly.  

\begin{figure*}
    \centering
    \includegraphics[width=0.55\textwidth]{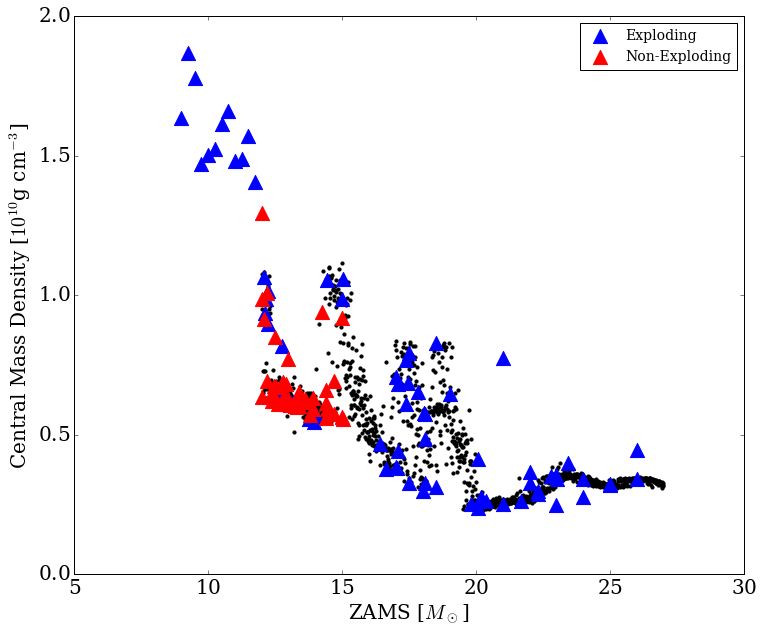}
    \caption{Initial central density versus ZAMS mass. The black dots are a comprehensive set of progenitor models found in the \citet{swbj16} and \citet{sukhbold2018}. Blue and red triangles are the 100 2D simulations in this work. Blue triangles are the exploding ones, while red triangles are non-exploding models. Most non-exploding models have a similar central density around $6\times10^9$ g cm$^{-3}$}
    \label{fig:rhoc-compactness}
\end{figure*}

We emphasize that detailed stellar evolution calculations \citep{swbj16,sukhbold2018} find that the initial structures of collapse depend in a ``chaotic" way on, among other things, shell mergers and the transition from radiative to convective core carbon burning (near $\sim$18 M$_{\odot}$?).  The different branches that emerge are seen clearly in Figure \ref{fig:he-compactness}, where almost all the failed explosions reside on one branch.  The larger significance of this clustering remains to be determined, but it seems to coincide with the model space in the \citet{sukhbold2018} collection that experiences two simultaneous oxygen burning shells.  In their model suite, above the associated ZAMS mass range ($\ge$15 M$_{\odot}$) one of these shells diminishes or disappears, leaving the other to dominate. This seems to result in a more substantial density jump at the ``residual" Si/O interface for those more massive progenitors. A similar branch dependence is seen in Figure \ref{fig:rhoc-compactness}, which renders the initial central mass density versus ZAMS mass. This plot too suggests an interesting relationship between branches in density structure that result from the consequences of the chaotic complications that attend core and shell burning of massive stars during their terminal stages. 

\begin{figure*}
    \centering
    \includegraphics[width=0.55\textwidth]{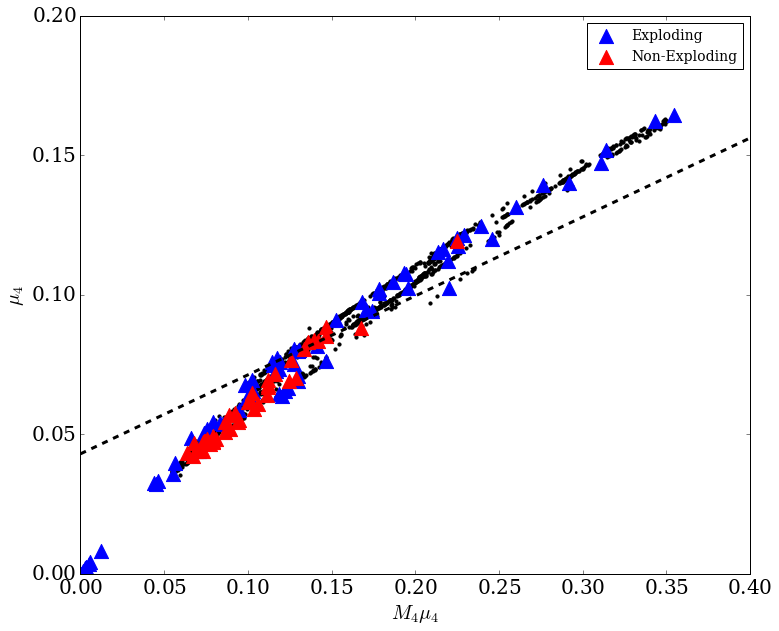}
    \caption{The Ertl condition plot. $M_4$ is the mass coordinate where the specific entropy first exceeds 4$k_b/$baryon, which roughly marks the position of the Si/O interface, and $\mu_4=\frac{0.3M_\odot\times1000\text{km}}{R(M_4+0.3M_\odot)-R(M_4)}$. The black dots are a comprehensive set of progenitor models found in the \citet{swbj16} and \citet{sukhbold2018}. Blue and red triangles are the 100 2D simulations in this work. Blue triangles are the exploding ones, while red triangles are non-exploding models. The dashed line is the w18.0 condition in \citet{2016ApJ...818..124E}. Almost all non-exploding models are below the Ertl et al. line, but the exploding models are both above and below the line.}
    \label{fig:ertl-compactness}
\end{figure*}

Finally, in Figure \ref{fig:ertl-compactness} we test the \citet{2016ApJ...818..124E}
condition against our 100-model suite. We see that though it finds most of the non-exploding models, it fails to find many of the exploding ones.

\section{Conclusions}
\label{conclude}

\begin{figure*}
    \centering
    \includegraphics[width=0.55\textwidth]{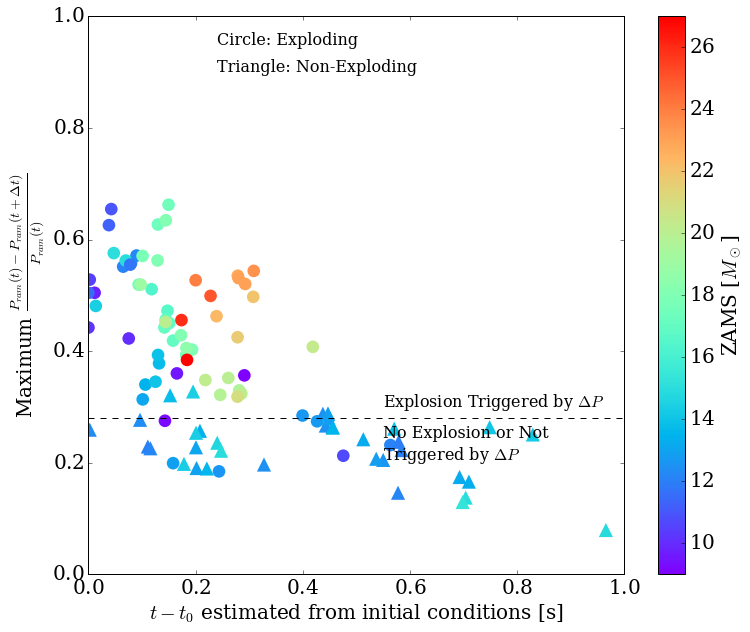}
    \caption{Same as Figure \ref{fig:pjump-all}, but for the 100 models we actually simulated in 2D for this study. The dashed line shows the simple explosion line from the right panel of Figure \ref{fig:pjump-ini}. Circular dots are the exploding models, while the triangles are non-exploding ones. Among 100 simulations, there are only $\sim$10 exceptions to our pressure jump criterion. Five of them do not explode, but are above the line (false positives), and six of them  explode, but are below the line (false negatives). Three false positive dots and two false negative dots are all very close to the criterion line, which means that 1) there might be a marginally exploding region, 2) the demarcation line can be slightly shifted and/or needn't be straight, or 3) there are secondary parameters of relevance when things are marginal. Other false negative dots can be explained by steep density profiles, in which the abrupt density jump is not strong, but the overall steepness of the density profile triggers the explosion. See text for a discussion.}
    \label{fig:pjump-100}
\end{figure*}

Calibrating with detailed 2D core-collapse supernova simulations, we derived a simple explosion condition based solely on the density profiles of the terminal cores provided by state-of-the-art stellar evolution calculations of progenitor massive stars.  This condition captures the vast majority of the behavior of the 100 2D models we performed subsequently to gauge its usefulness. The goal was to be able to predict, without resort to detailed simulation, the explodability of a given massive star.  Under the assumption that our 2D multi-group radiation/hydrodynamic simulations using the code F{\sc{ornax}} yield reliable results, we find that the simple maximum fractional ram pressure jump discriminant we defined works well ($\sim 90$\% of the time). We emphasize that this method employs simple analytic prescriptions for the various inputs needed that are based solely upon a given initial mass density profile inherited from detailed stellar model calculations. The maximum ram pressure jump generally occurs at the time of accretion of the Si/O interface, but not always. We note that the false positives are all close to the horizontal demarcation line (see Figure \ref{fig:pjump-100}), suggesting that this line can be altered in shape only slightly to include them or that there may be other ``second-order" criteria that come into play.  These could be the initial thermal and/or Y$_e$ profiles.  The latter have a slight influence over the timing of accretion not factored into our crude analytic model for the mass accretion rate evolution through the shock, incorporating as it does only the initial density profile.  

There seem to be two types of false negatives. The first is the models on Figure \ref{fig:pjump-100} with $\frac{\Delta P_{\rm ram}}{P_{\rm ram}}\sim0.2$.  These models have masses below $\sim$12.5 M$_{\odot}$ and have steep initial density profiles and low compactness.  Such models don't need a density discontinuity (such as at a pronounced Si/O interface) to explode. The second class of false negatives involves models that do not have steep density profiles and explode later after bounce. They are much closer to the horizontal line. Though in drawing our sample of progenitors we were initially guided by the Salpeter initial mass function, we can not claim to have done so rigorously.  Hence, the fraction of models we see explode in this investigation should not be confused with the fraction of massive stars in Nature above $\sim$8 M$_{\odot}$ that do explode.  This fraction, and the fraction that result in black holes, has yet to be determined.  However, should our explodability condition have some validity, this fraction can easily be calculated in the context of a definitive suite of progenitor models spanning the requisite range of massive-star masses. 

It should be mentioned that we drew our initial models from both the \citet{swbj16} and \citet{sukhbold2018} compilations.  All the models with $\le$12 M$_{\odot}$ came from the former, which sampled the associated mass range far more sparsely than did \citet{sukhbold2018} for masses above 12 M$_{\odot}$.  This is reflected in the greater number of studied models above 12 M$_{\odot}$. Also, we did not include progenitors with ZAMS masses above 27 M$_{\odot}$.  We suspect that many models in this higher mass range do not explode, or may explode initially, but experience fallback later upon encountering an outer mantle with significant binding energy that the blast can not fully overcome.  The result would be late-time fallback, but there might still be some mass ejection to infinity with a reduced asymptotic explosion energy \citep{2018ApJ...852L..19C}. This intriguing possibility remains to be demonstrated, but would suggest that even black hole formation might be associated with some sort of (supernova) explosion.

There are many caveats that deserve mention.  Our results are conditioned on the fidelity with which our current implementation of F{\sc{ornax}} adheres to Nature.  There remain issues concerning the neutrino-matter interaction, the nuclear equation of state, the possible effects of neutrino oscillations, and the accuracy of the numerical algorithms employed.  Furthermore, we have noted in the past that the spatial resolution of the computational grid can be a factor in the outcome, with higher resolution models exploding when lower resolution models don't \citep{nagakura2019b}.  We have endeavored in this investigation to employ adequate resolution, but this remains to be definitely demonstrated. Moreover, we are presuming that models that explode in 2D also explode in 3D, and vice versa. This has generally been our experience. We note that in 3D the horizontal explosion/no-explosion demarcation line may need to be slightly adjusted (see Figures \ref{fig:pjump-sim} and \ref{fig:pjump-ini}). However, determining whether this is necessary must await a large study using a corresponding number of detailed 3D simulations. In addition, we have not addressed the effect of the initial seed perturbations \citep{2013ApJ...778L...7C,muller2017,burrows2018} in kick-starting the turbulence behind the stalled shock, nor the fact that chaotic flow will by its nature lead to stochastic outcomes \footnote{though whether only in detail, or qualitatively, has not been determined}.  It is also the case that even state-of-the-art progenitor modeling is still in flux; the mapping between progenitor structure and ZAMS mass, for instance, has not converged. Finally, we have not addressed in this study the possible roles of \tianshu{the equation of state, rotation or magnetic fields \citep{burrows2007_mag,2013ApJ...764...99S,2013ApJ...765...29C,Mosta2014,PhysRevLett.124.092701,2020ApJ...896..102K,Obergaulinger2020,Aloy2021,Obergaulinger2021,2021ApJ...906..128K}}. 

Nevertheless, the straightforward explodability condition we have obtained has been calibrated with a large collection of state-of-the-art 2D simulations and is simple to implement.  It demonstrates fidelity to our more detailed simulations $\sim$90\% of the time and shows promise that it might be further generalized while still employing data from only the unstable Chandrasekhar progenitors core-collapse modelers inherit. We suggest it is a more credible and robust predictor than can currently be found in the literature, but plan to refine it in numerous ways in the future. 

\section*{Data Availability}
The data underlying this article will be shared on reasonable request to the corresponding author.

\section*{Acknowledgements}

We thank Chris White, David Radice, and Hiroki Nagakura for insights and advice during the germination of this project. We acknowledge support from the U.~S.\ Department of Energy Office of Science and the Office of Advanced Scientific Computing Research via the Scientific Discovery through Advanced Computing (SciDAC4) program and Grant DE-SC0018297 (subaward 00009650) and support from the U.~S.\ National Science Foundation (NSF) under Grants AST-1714267 and PHY-1804048 (the latter via the Max-Planck/Princeton Center (MPPC) for Plasma Physics). The three-dimensional simulations were performed on Blue Waters under the sustained-petascale computing project, which was supported by the National Science Foundation (awards OCI-0725070 and ACI-1238993) and the state of Illinois. Blue Waters was a joint effort of the University of Illinois at Urbana--Champaign and its National Center for Supercomputing Applications. We also acknowledge access to the Frontera cluster (under awards AST20020 and AST21003), and this research is part of the Frontera computing project at the Texas Advanced Computing Center \citep{Stanzione2020}. Frontera is made possible by NSF award OAC-1818253. Additionally, a generous award of computer time was provided by the INCITE program, enabling this research to use resources of the Argonne Leadership Computing Facility, a DOE Office of Science User Facility supported under Contract DE-AC02-06CH11357. Finally, the authors acknowledge computational resources provided by the high-performance computer center at Princeton University, which is jointly supported by the Princeton Institute for Computational Science and Engineering (PICSciE) and the Princeton University Office of Information Technology, and our continuing allocation at the National Energy Research Scientific Computing Center (NERSC), which is supported by the Office of Science of the U.~S.\ Department of Energy under contract DE-AC03-76SF00098.

\clearpage

\bibliographystyle{mnras}
\bibliography{References}

\bsp	
\label{lastpage}
\end{document}